\documentclass{emulateapj}
\usepackage{subfigure}
\usepackage{amsmath} 
\usepackage{graphicx}
\usepackage{textcomp}
\usepackage{color}
\usepackage{float}
\pdfoutput=1
\usepackage{ifpdf}

\begin{document}
\title{A multi-wavelength polarimetric study of the blazar CTA102 during a gamma-ray flare in 2012}
\author{Carolina Casadio\altaffilmark{1}, Jos\'e L. G\'omez\altaffilmark{1}, Svetlana G. Jorstad\altaffilmark{2,3}, 
Alan P. Marscher\altaffilmark{2}, Valeri M. Larionov\altaffilmark{3,6}, Paul S. Smith\altaffilmark{4}, Mark A. Gurwell\altaffilmark{5}, Anne L\"ahteenm\"aki\altaffilmark{6,7}, Iv\'an Agudo\altaffilmark{1}, Sol N. Molina\altaffilmark{1}, Vishal Bala\altaffilmark{2}, Manasvita Joshi\altaffilmark{2}, Brian Taylor\altaffilmark{2}, Karen E. Williamson\altaffilmark{2}, 
Arkady A. Arkharov\altaffilmark{8}, Dmitry A. Blinov\altaffilmark{9,3}, George A. Borman\altaffilmark{10}, Andrea Di Paola\altaffilmark{11}, Tatiana S. Grishina\altaffilmark{3}, Vladimir A. Hagen-Thorn\altaffilmark{3}, 
Ryosuke Itoh\altaffilmark{12}, Evgenia N. Kopatskaya\altaffilmark{3}, Elena G. Larionova\altaffilmark{3}, Liudmila V. Larionova\altaffilmark{3}, Daria A. Morozova\altaffilmark{3}, Elizaveta Rastorgueva-Foi\altaffilmark{6,13}, 
Sergey G. Sergeev\altaffilmark{10}, Merja Tornikoski\altaffilmark{6}, Ivan S. Troitsky\altaffilmark{3}, Clemens Thum\altaffilmark{14}, Helmut Wiesemeyer\altaffilmark{15}}

\altaffiltext{1}{Instituto de Astrof\'{\i}sica de Andaluc\'{\i}a, CSIC, Apartado 3004, 18080, Granada, Spain}
\altaffiltext{2}{Institute for Astrophysical Research, Boston University, 725 Commonwealth Avenue, Boston, MA 02215}
\altaffiltext{3}{Astronomical Institute, St. Petersburg State University, Universitetskij Pr. 28, Petrodvorets, 198504 St. Petersburg, Russia}
\altaffiltext{4}{Steward Observatory, University of Arizona, Tucson, AZ  85716  USA}
\altaffiltext{5}{Harvard-Smithsonian Center for Astrophysics, Cambridge, MA 02138, USA}
\altaffiltext{6}{Aalto University Mets\"ahovi Radio Observatory, Mets\"ahovintie 114, 02540 Kylm\"al\"a, Finland}
\altaffiltext{7}{Aalto University Department of Radio Science and Engineering, P.O. BOX 13000, FI-00076 AALTO, Finland}
\altaffiltext{8}{Pulkovo Observatory, St. Petersburg, Russia}
\altaffiltext{9}{University of Crete, Heraklion, Greece}
\altaffiltext{10}{Crimean Astrophysical Observatory, Russia}
\altaffiltext{11}{INAF, Osservatorio Astronomico di Roma, Italy}
\altaffiltext{12}{Department of Physical Sciences, Hiroshima University, Higashi-Hiroshima, Hiroshima 739-8526, Japan}
\altaffiltext{13}{School of Maths and Physics, University of Tasmania, Australia, Private Bag 37, Hobart TAS 7001}
\altaffiltext{14}{Instituto de Radio Astronom\'{i}a Milim\'{e}trica, Granada, Spain}
\altaffiltext{15}{Max-Planck-Institut f\"ur Radioastronomie, Bonn, Germany}

\shorttitle{}
\shortauthors{}
\begin{abstract}   
  We perform a multi-wavelength polarimetric study of the quasar CTA~102 during an extraordinarily bright $\gamma$-ray outburst detected by the {\it Fermi} Large Area Telescope in September-October 2012 when the source reached a flux of F$_{>100~\mathrm{MeV}} =5.2\pm0.4\times10^{-6}$ photons cm$^{-2}$ s$^{-1}$. 
  At the same time the source displayed an unprecedented optical and NIR outburst. We study the evolution of the parsec-scale jet with ultra-high angular resolution through a sequence of 80 total and polarized intensity Very Long Baseline Array images at 43 GHz, covering the observing period from June 2007 to June 2014. We find that the $\gamma$-ray outburst is coincident with flares at all the other frequencies and is related to the passage of a new superluminal knot through the radio core. The powerful $\gamma$-ray emission is associated with a change in direction of the jet, which became oriented more closely to our line of sight ($\theta\sim$1.2$^{\circ}$) during the ejection of the knot and the $\gamma$-ray outburst. 
  During the flare, the optical polarized emission displays intra-day variability and a clear clockwise rotation of EVPAs, which we associate with the path followed by the knot as it moves along helical magnetic field lines, although a random walk of the EVPA caused by a turbulent magnetic field cannot be ruled out. We locate the $\gamma$-ray outburst a short distance downstream of the radio core, parsecs from the black hole. This suggests that synchrotron self-Compton scattering of near-infrared to ultraviolet photons is the probable mechanism for the $\gamma$-ray production.

\end{abstract}  


\section{Introduction}
The Blazar CTA~102 (B2230+114) is classified as a highly polarized quasar (HPQ), characterized by optical polarization exceeding 3\%~\citep{Moore:1981fk}. Its high variability at optical wavelengths~\citep{Pica:1988kx, Osterman-Meyer:2009uq} and its spectral properties identify it also as an optically violent variable (OVV) quasar~\citep{Maraschi:1986kx}. Microvariability of CTA~102 at optical wavelengths has been investigated by \cite{Osterman-Meyer:2009uq}, who found that faster variability is associated with higher flux states.

  The variability in this source occurs at other frequencies as well: flares at cm and mm wavelengths have been registered in the past, as well as an X-ray flare detected by the {\it Rossi X-ray Timing Explorer} in late 2005 \citep{Osterman-Meyer:2009uq}. A radio flare in 1997 was related to the ejection of a new knot from the core \citep{Savolainen:2002ys, Rantakyro:2003fk, Jorstad:2005fk}, and another, in 2006, was connected with both the ejection of a new superluminal feature and the interaction between this component and a recollimation shock at 0.1 mas~\citep{Fromm:2011zr}.

  The radio morphology on arcsecond scales (from images with the Very Large Array at 6 and 2 cm) reveals a central core with two weak components located at opposite sides~\citep{Spencer:1989vn, Stanghellini:1998ys}. At higher angular resolution, CTA~102 has been regularly observed since 1995 within the VLBA 2cm-Survey \citep[e.g.][]{Zensus:2002uq} and its successor, the MOJAVE program \citep[e.g.,][]{Lister:2009kx}. MOJAVE images show that the jet in CTA~102 extends toward the southeast up to $\sim$15 mas from the core, which corresponds to a de-projected distance of $\sim$ 2.7 kpc using the estimated viewing angle of 2.6$^{\circ}$ obtained by \cite{Jorstad:2005fk}.
  
  Kinematic studies of the MOJAVE data report apparent velocities between 1.39$c$ and 8.64$c$ \citep{Lister:2013vn}. Higher apparent speeds, up to ${\beta_{app}}\sim$18$c$, have been reported in higher-resolution Very Long Baseline Array (VLBA) observations at 43 GHz by \cite{Jorstad:2001fk,Jorstad:2005fk}. Apart from superluminal features, the jet of CTA~102 also displays standing features: two sationary components, A1 and C, have been observed at a distance of $\sim$0.1 and 2 mas from the core, respectively \citep{Jorstad:2001fk,Jorstad:2005fk}, and interpreted as recollimation shocks \citep{Fromm:2013uq}.

  Recent MOJAVE polarimetric multifrequency observations~\citep{Hovatta:2012fk} reveal a rotation measure gradient across the jet width at 7 mas from the core, which suggests the presence of a helical magnetic field in the jet. A similar result is reported in~\cite{Fromm:2013uq}, where the different observing frequencies reveal bends and helical structures in many parts of the jet.

  CTA~102 was detected by the {\it Fermi\/} Gamma-ray Space Telescope in the first {\it Large Area Telescope\/} (LAT) catalog with a flux (E$>100~\mathrm{MeV}$) of $14.70\pm0.97\times10^{-8}$ photons cm$^{-2}$ s$^{-1}$~\citep{Abdo:2010fk}, and confirmed in the second catalog~\citep{Ackermann:2011uq}. In late 2012 the blazar exhibited an extraordinarily bright $\gamma$-ray outburst detected by the LAT, 
reaching a flux of $5.17\pm0.44\times10^{-6}$ photons cm$^{-2}$ $s^{-1}$ between 0.1 and 200 GeV. During the 2012 event, an unprecedented optical and NIR outburst was observed, as reported by \cite{Larionov:2012eu}  and~\cite{Carrasco:2012fp}, respectively.

  In this paper we perform a radio to $\gamma$-ray multi-wavelength analysis in order to study the flaring activity of CTA~102 during the 2012 event. In Section 2, we present the multi-wavelength data set collected for the analysis, and we describe the methods used to reduce the data. In Section 3 we describe the characteristics of the emission at the different frequencies during the 2012 flare event. In Section 4 we study the kinematics and the flux density variability of the parsec scale jet. In Section 5 we perform the discrete cross-correlation analysis between light-curves. In Section 6 we analyze the polarized emission of the source at mm and optical wavebands. We present our discussions and conclusions in Sections 7 and 8.

  We adopt the cosmological values from the most recent {\it Planck} satellite results \citep{Planck-Collaboration:2014}: $\Omega_{m}$= 0.3, $\Omega_{\Lambda}$= 0.7, and $H_{0}$ = 68 km s$^{-1}$ Mpc$^{-1}$. With these values, at the redshift of CTA~102 ($z$=1.037), 1 mas corresponds to a linear distance of 8.31 pc, and a proper motion of 1 mas yr$^{-1}$ corresponds to an apparent speed of 55.2$c$.


\section{Observations and Data Reduction}

  We are interested in studying the multi-spectral behavior of CTA~102 during the $\gamma$-ray flare that occurred between 2012 September and October. For this, we have collected data from millimeter to $\gamma$-ray wavelengths, extending our study from 2004 June to 2014 June. In particular, the {\it Fermi} data extend from the start of the mission (2008 August) to 2013 September, X-ray and UV data cover the observing period from 2009 August to 2013 June, optical and NIR data from 2004 June to 2013 October, and the combined radio light curves cover the entire period from 2004 June to 2014 June.

  In the mm-wave range, we use data at (1) 350~GHz (0.85~mm) and 230~GHz (1.3~mm), obtained with the Submillimeter Array (SMA) at Mauna Kea, Hawaii; (2) 230~GHz (1.3~mm) and 86.24~GHz (3.5~mm) with the 30 m Telescope of Institut de Radioastronomie Millim\'etrique (IRAM) at the Pico Veleta Observatory (Spain); (3) 43~GHz (7~mm) with the Very Long Baseline Array (VLBA); and (5) 37~GHz (8~mm) with the 13.7 m Telescope at Mets\"ahovi Radio Observatory of Aalto University (Finland).

  Near-infrared photometric data (JHK filters) were obtained at the Perkins Telescope at Lowell Observatory (Flagstaff, AZ) using the MIMIR instrument \citep{Clemens:2007fk} and at the 1.1 m Telescope of the Main Astronomical Observatory of the Russian Academy of Sciences located at Campo Imperatore, Italy \citep[see][for details]{Hagen-Thorn:2008uq}. 

  We have collected optical photometric data in UBVRI bands from numerous telescopes: (1) the 2.2~m Telescope of Calar Alto Observatory (Almer\'ia, Spain)\footnote{Observations performed under the MAPCAT (Monitoring AGN with the Calar Alto Telescopes), see \cite{Agudo:2012fk}}; (2) the 2~m Liverpool Telescope of the Observatorio del Roque de Los Muchachos (Canary Island, Spain); (3) the 1.83~m Perkins Telescope of Lowell Observatory (Flagstaff, AZ); (4) the 1.54~m and 2.3~m telescopes of Steward Observatory (Mt. Bigelow and Kitt Peak, AZ)\footnote{Data taken from the Steward Observatory monitoring project, see \cite{Smith:2009fk}}; (5) the 40-cm LX-200 Telescope of St. Petersburg State University (St. Petersburg, Russia); 
  (6) the 70~cm AZT-8 Telescope of the Crimean Astrophysical Observatory (Nauchnij, Ukraine); (7) the 1.5~m Kanata Telescope in Higashi-Hiroshima Observatory (Japan)\footnote{Data published in \cite{Itoh:2013uq}}; and (8) the Ultraviolet and Optical Telescope (UVOT) on board the {\it Swift} satellite. 
  Optical data are in part also in polarimetric mode, mostly in R band, except for items (4) and (5) listed above \citep[see][respectively, for details]{Schmidt:1992kx,Hagen-Thorn:2008uq}.
  In the UV range we use UVOT data from {\it Swift} in the three available filters: UVW2 (2030 {\AA}), UVM2 (2231 {\AA}) and UVW1 (2634 {\AA}). We have also obtained X-ray data in the energy range 0.3--10 keV from the X-Ray Telescope (XRT) on board of {\it Swift} satellite.

  At the highest photon energies considered here, we have analyzed $\gamma$-ray data from the Large Area Telescope (LAT) of the {\it Fermi} Gamma-ray Space Telescope.


\subsection {\it$\gamma$-ray Data Analysis}

  We have analyzed the $\gamma$-ray data of the field containing CTA~102 from the {\it Fermi} LAT from 2008 August to 2013 September, producing a light curve between 0.1 and 200 GeV with an integration time of 1 day. We used the Fermi Science Tools version v9r33p0 and instrument response function P7REP$\_$SOURCE$\_$V15, considering data inside a region of interest (ROI) of 15$^{\circ}$ radius centered on CTA~102. We followed the unbinned likelihood procedure as described in web page http://fermi.gsfc.nasa.gov/ssc/data/analysis/scitools/.

\begin{figure}[htpb]
\centering
\includegraphics[width=0.46\textwidth]{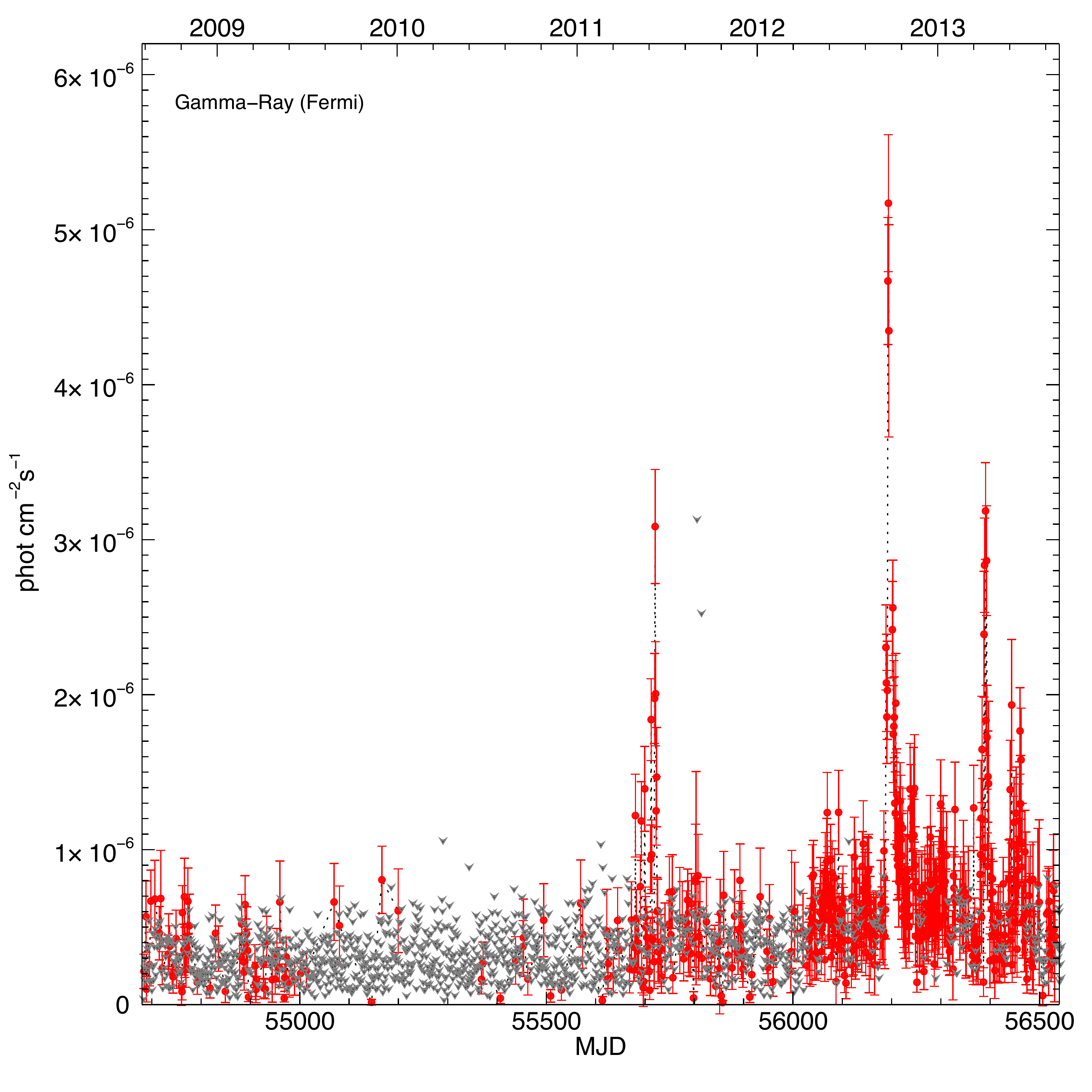}
\caption{{\it Fermi} LAT $\gamma$-ray light curve between 0.1 and 200 GeV, with an integration time of 1 day. Red points represent the detections (TS$>$10) and gray arrows correspond to upper limits when the source is not detected (TS$<$10).}
\label{fig:CTA102_gamma_lc_1day}
\end{figure}

  The procedure starts with the selection of good data and time intervals through the tasks {\it gtselect} and {\it gtmktime}, and follows with the creation of an exposure map for each day (tasks {\it gtltcube}, {\it gtexpmap}) and the modeling of data through a maximum-likelihood method ({\it gtlike}). In this last step, we used a model that includes CTA~102 and 42 other point sources inside the ROI, according to the second {\it Fermi} Gamma-ray Catalog, \citep[2FGL;][]{Ackermann:2011uq}. 
  The model also takes into account the diffuse emission from our Galaxy ({\it gll$\_$iem$\_$v05.fit}) and the extragalactic $\gamma$-ray background ({\it iso$\_$source$\_$v05.txt}). We searched for the flux normalization of CTA~102 by fixing the spectral index of the other point sources while varying both the flux and spectral index of our target. The $\gamma$-ray spectrum of CTA~102 was modeled with a log parabola curve corresponding to the spectral model given in the 2FGL catalog. 
  We considered a successful detection when the test statistic $TS \geq 10$, which corresponds to a signal-to-noise ratio $\gtrsim 3$-$\sigma$~\citep{Nolan:2012fk}.


\begin{figure*}[t]
\centering
\includegraphics[width=0.85\textwidth]{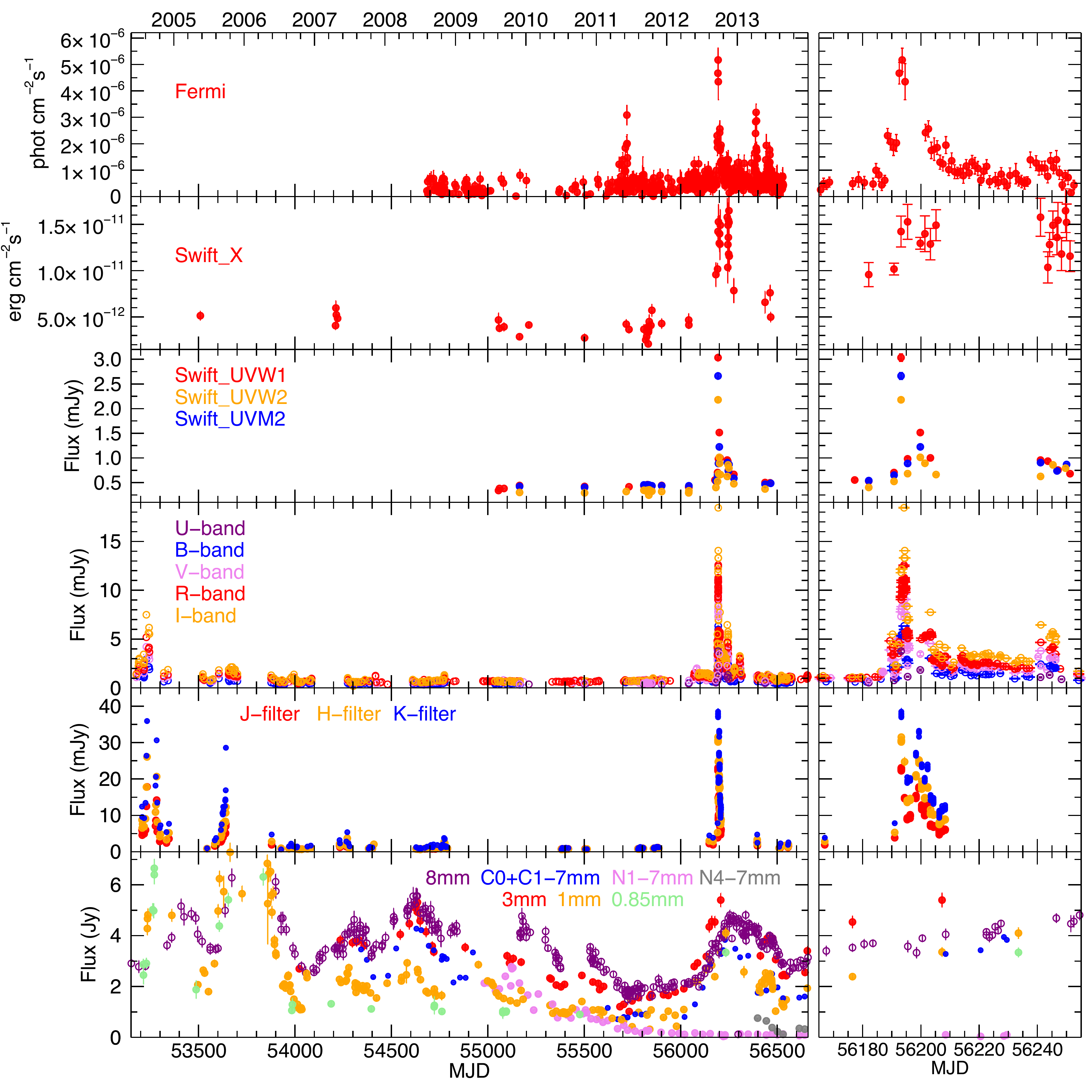}
\caption{Light curves of CTA~102 from $\gamma$-ray to millimeter wavelengths. From top to bottom: $\gamma$-ray, X-ray, UV, optical, NIR and millimeter-wave data. {\it Left panel:} Data from 2004 May to 2014 January. {\it Right panel:} Expanded view during the $\gamma$-ray outburst between 2012 August and November.}
\label{fig:CTA102_mwl_lc}
\end{figure*}

\subsection {\it X-ray, UV, Optical and Near Infrared} 

  We collected X-ray and UV data from 2009 August to 2013 June from the {\it Swift} archive. The X-ray data in the energy range 0.3--10 keV were calibrated following the procedure described in \cite{Williamson:2014kx}. The UVOT data reduction was performed via the UVOTSOURCE tool, with a sigma value of 5 adopted to compute the background limit.
  Optical and NIR data were reduced and calibrated following the procedures outlined in \cite{Jorstad:2010fk}.

  All of the magnitudes of the optical and NIR data have been corrected for Galactic extinction with values reported in the NASA Extragalactic Database (NED)\footnote{http://ned.ipac.caltech.edu/} for each filter \citep{Schlafly:2011kx}. For the UV data, we obtained the absolute extinction value at each wavelength A($\lambda$) from equation (1) in \cite{Cardelli:1989vn}. After the correction we transformed magnitudes into fluxes using the formula reported in \cite{Mead:1990kx} and \cite{Poole2008}.


\subsection {\it Photo-Polarimetric Millimeter VLBA and Single Dish Data}

  Multi-epoch very long baseline interferometer (VLBI) images provides ultra-high angular resolution that can be used to determine the location in the jet where flaring activity occurs. We therefore have collected data from the VLBA-BU-BLAZAR program\footnote{http://www.bu.edu/blazars/research.html}, which consists of monthly monitoring of $\gamma$-ray bright blazars with the VLBA at 43 GHz (7~mm). The dataset consists of 80 total and polarized intensity images from 2007 June to 2014 June. 
  We restore the images with a common convolving beam of 0.4$\times$0.2 milliarcseconds (mas). Since the resolution of the longest baselines of the VLBA is less than half of these dimensions, we employ model fitting to define and determine the parameters of the very fine-scale structure. The data reduction was performed with a combination of the Astronomical Image Processing System ({\tt AIPS}) and the Differential Mapping software ({\tt Difmap}), as described in \cite{Jorstad:2005fk}. 
  The electric vector position angle (EVPA) calibration follows the procedure discussed in~\cite{Jorstad:2005fk}, which combines the comparison between VLA and VLBA integrated EVPA values at those epochs for which VLA data are available with the method of \cite{Gomez:2002kx} that utilizes the stability of the instrumental polarization (D-terms).
  
  The IRAM 30m Telescope's total flux and polarimetric data in this paper were acquired under the POLAMI (Polarimetric AGN Monitoring at the IRAM 30m Telescope) program (see Agudo et al. MNRAS in prep.) 
  and reduced and calibrated following the procedures introduced in \cite{Agudo:2006uq, Agudo:2010kx, 
  Agudo:2014vn}.
  
  The Submillimiter Array data of CTA~102 came from an ongoing monitoring program at the SMA to determine the fluxes of compact extragalactic radio sources that can be used as calibrators at mm wavelengths \citep{Gurwell:2007ys}. 
  Observations of available potential calibrators are from time to time observed for 3 to 5 minutes, and the measured source signal strength calibrated against known standards, typically solar system objects (Titan, Uranus, Neptune, or Callisto).  Data from this program are updated regularly and are available at the SMA website.\footnote{http://sma1.sma.hawaii.edu/callist/callist.html}
  

\section{Multi-wavelength Outburst} 

Figure \ref{fig:CTA102_gamma_lc_1day} displays the $\gamma$-ray light curve of CTA~102 in the energy range 0.1--200 GeV during the period of major activity (2011 June--2013 April) obtained with an integration time of 1 day. Following \cite{Jorstad:2013uq} we can define a $\gamma$-ray outburst as a  period when the flux exceeds a threshold of $2\times10^{-6}$ photons cm$^{-2}$ s$^{-1}$. Although this is an arbitrary limit, it conforms to a visual inspection of the $\gamma$-ray light curve of CTA~102.

  The first outburst takes place in 2011 June (MJD 55719-55721), when the source displays a one-day peak flux of $3.1\pm 0.37\times10^{-6}$ photons cm$^{-2}$ s$^{-1}$. The second, brightest outburst occurs at the end of 2012 September (2012.73), when the source remains above $2\times10^{-6}$ photons cm$^{-2}$ s$^{-1}$ for 14 days (from MJD 56188 to 56202), reaching a peak of $5.2\pm0.4\times10^{-6}$ photons cm$^{-2}$ s$^{-1}$ on MJD 56193. During this outburst the $\gamma$-ray flux increases by a factor of ten in just 6 days. The third flare occurs in April 2013 (MJD 56387--56394) and lasts 8 days. On this occasion (MJD 56392), the blazar reaches a peak of $2.9\pm0.4\times10^{-6}$ photons cm$^{-2}$ s$^{-1}$.

  We compare the daily $\gamma$-ray light-curve with the X-ray, UV, optical, NIR, and radio light-curves in Figure~\ref{fig:CTA102_mwl_lc}. Table \ref{tabMulti} lists the multi-wavelength data used in our analysis. The brightest $\gamma$-ray outburst, in 2012 September, is accompanied by similarly bright flares at all of the other wave bands. This is, however, not the case for the other two orphan $\gamma$-ray flares ($\gamma$-ray outbursts with no correspondence at any of the other observing bands), with the exception of a mm-wave flare that follows the third $\gamma$-ray flare in 2013 April.
  
\begin{figure}[htpb]
\centering
\includegraphics[width=0.48\textwidth]{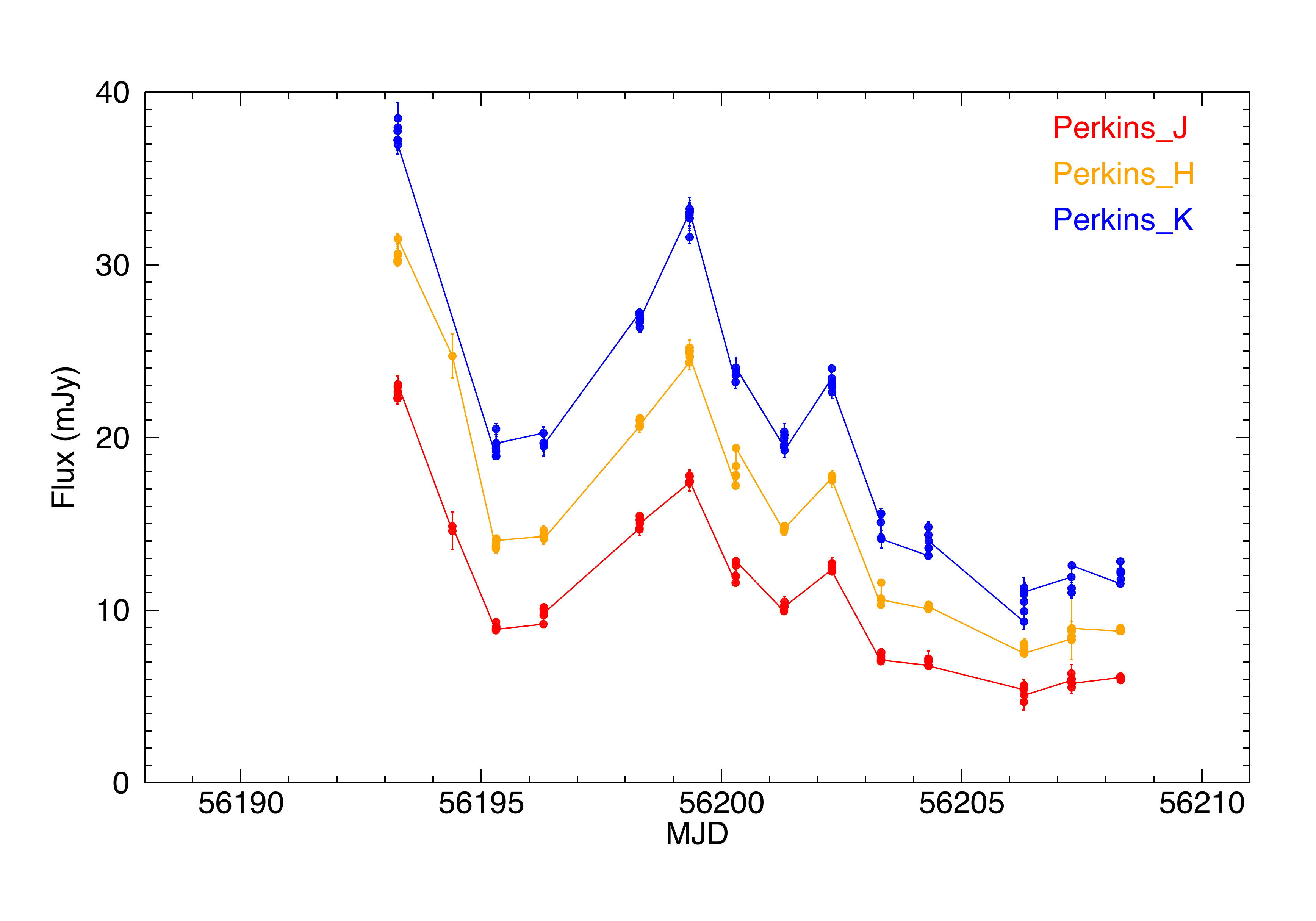}
\caption{An expanded view of the NIR light-curve in the J, H, and K bands during the period of the main $\gamma$-ray outburst.}
\label{fig:CTA102_NIR}
\end{figure}

  Analyzing the multi-wavelength flare in 2012, we observe that the X-ray light curve contains a double-peak structure, where the first peak is almost coincident with the $\gamma$-ray outburst and the second peak occurs $\sim$50 days later. The limited sampling of the X-rays prevents a deeper analysis of the overall X-ray behavior associated with this flare. The UV and optical bands exhibit a rapid and pronounced increase in the light curves coinciding with the $\gamma$-ray flare. In the optical light curve, we distinguish a secondary, weaker flare after $\sim$50 days, close to the second X-ray peak, as well as a third, smaller outburst that occurs $\sim$70 days after the second peak. In the UV it is also possible to distinguish a secondary flare delayed by $\sim$50 days with respect to the main flare, but the sampling of the data is insufficient to specify the behavior in more detail. In the NIR light curve, we observe a large flare coincident with the $\gamma$-ray flare, but there is no further sampling after this. A detailed analysis of the NIR flare shown in Figure~\ref{fig:CTA102_NIR} reveals that the event consists of three sub-flares covering almost the entire period of high $\gamma$-ray flux from MJD 56193 to 56202.

\begin{table}[htbp]
  \begin{center}  
  \caption{Multi wavelength Data}  
   \begin{tabular}{ccc} 
     \hline
     \hline
$\gamma$-ray data\\
\hline
Epoch & Flux &  Energy band\\
(MJD) & (phot/cm$^{2}$/sec) & (GeV)\\
\hline 
 54684.2 &  2.15e-07$\pm$1.11e-07 &  0.1-200\\ 
 54688.2 &  5.68e-07$\pm$2.30e-07 &  0.1-200\\ 
 54689.2 & 10.00e-08$\pm$8.30e-08 &  0.1-200\\  
  .\\ 
  .\\ 
\hline 
X-ray data \\
\hline
Epoch & Flux &  Energy band\\
(MJD) & (erg/cm$^{2}$/sec) & (KeV)\\
\hline 
 53509.4 &  5.14e-12$\pm$4.74e-13 &  0.3-10\\
 54210.2 &  4.07e-12$\pm$4.13e-13 &  0.3-10\\
 54212.1 &  5.97e-12$\pm$7.67e-13 &  0.3-10\\   
\hline\\
\multicolumn{3}{c}{\footnotesize (This table is available in its entirety in machine-readable form.)}
\end{tabular} 
  \label{tabMulti}
  \end{center}
\end{table}

  The radio light curve also exhibits an increase in flux density during the 2012 $\gamma$-ray outburst, but with a much longer time scale, lasting $\sim$200 days. The 1 mm light curve and 7 mm light curves of features C0 and C1 peak on $\sim$ MJD 56230, about 1 month after the $\gamma$-ray flare. The 3 mm light curve follows a similar trend, starting to increase on $\sim$ MJD 56000. Our limited time sampling between MJD 56208 and 56412 shows a peak on MJD 56207, very close to the $\gamma$-ray flare, although we cannot rule out the possibility that the actual peak is closer to that at 1 mm.
  

\begin{figure*}[htpb]
\centering
\includegraphics[width=1.6\textwidth]{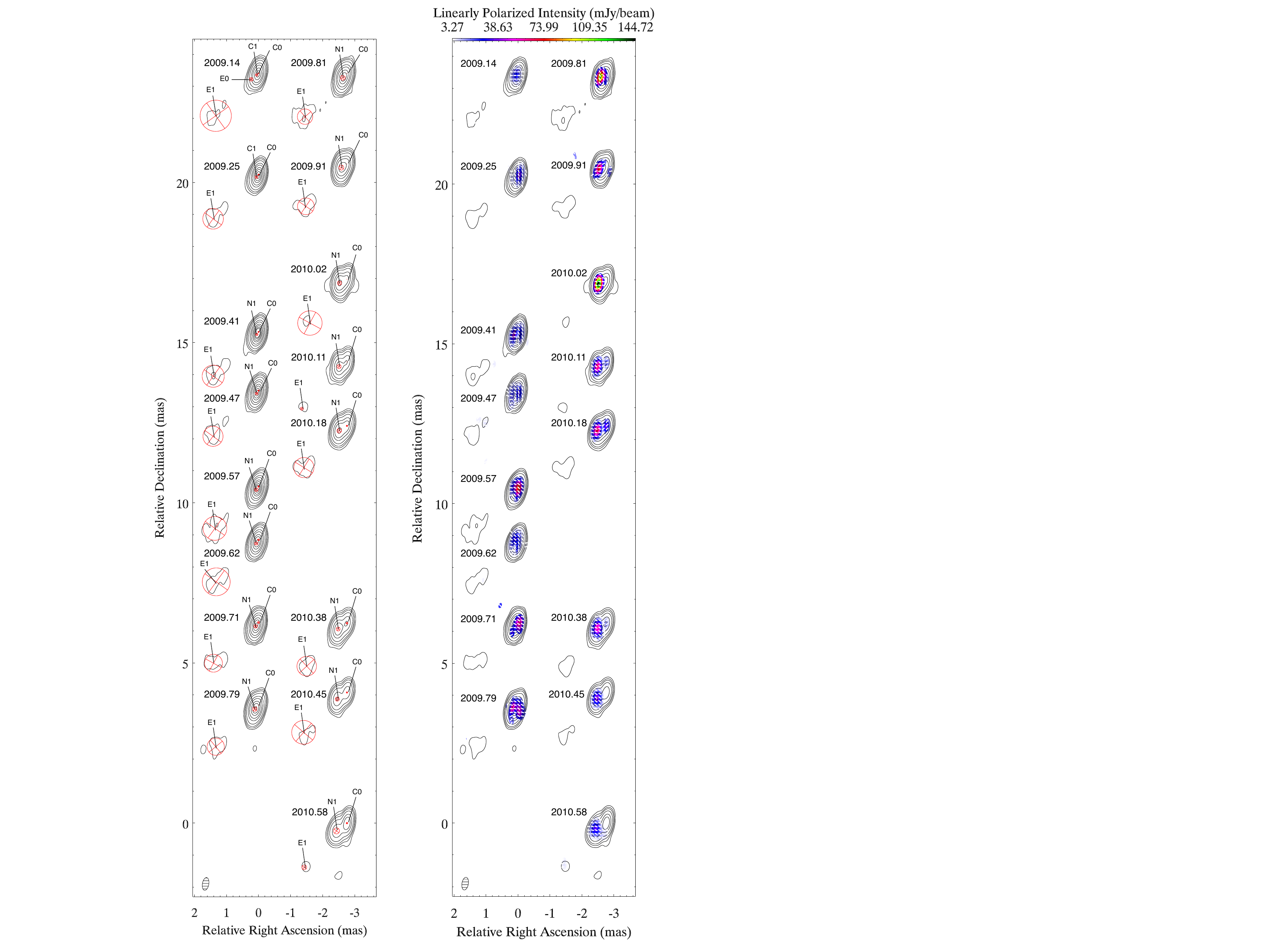}
\caption{Sequence of 43 GHz VLBA images displaying epochs from 2009 February to 2010 August, when we observe the appearance of component N1. The images are restored with a common beam of 0.4$\times$0.2 mas at $-$10$^{\circ}$ and are separated by a distance proportional to the time elapsed between observing epochs. Left panel: contours (total intensity) are traced at 0.003, 0.008, 0.04, 0.1, 0.3, 0.6, 1.2, 1.8, 2.5 and 3.0 Jy/beam and $I_{peak}$=4.2 Jy/beam. Red circles represent modelfit components. Right panel: same contours (total intensity) as in left panel plus colors that represent linearly polarized intensity and white sticks symbolizing linear polarization angle.}
\label{fig:CTA102_28}
\end{figure*}

\begin{figure*}[htbp]
\centering
\includegraphics[width=0.80\textwidth]{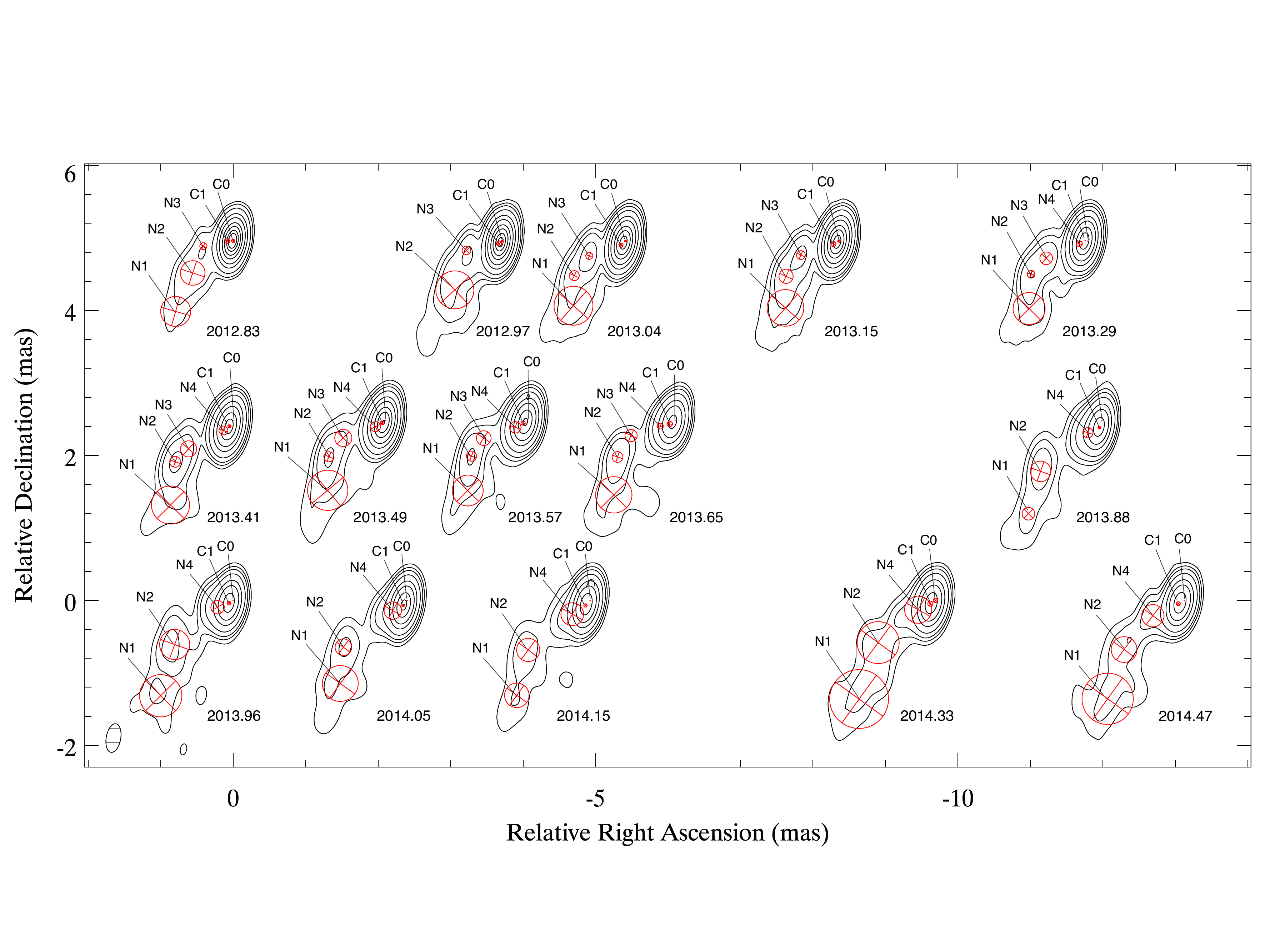}
\caption{Sequence of total intensity 43 GHz VLBA images from 2012 October to 2014 June, covering the epochs from which we start observing component N4. Peak intensity is $I_{peak}$=3.6 Jy/beam and contours are traced at 0.003, 0.008, 0.04, 0.1, 0.3, 0.6, 1.2, 1.8, 2.5 and 3.0 Jy/beam.}
\label{fig:CTA102_bw}
\end{figure*}

\begin{figure*}[htbp]
\centering
\includegraphics[width=0.80\textwidth]{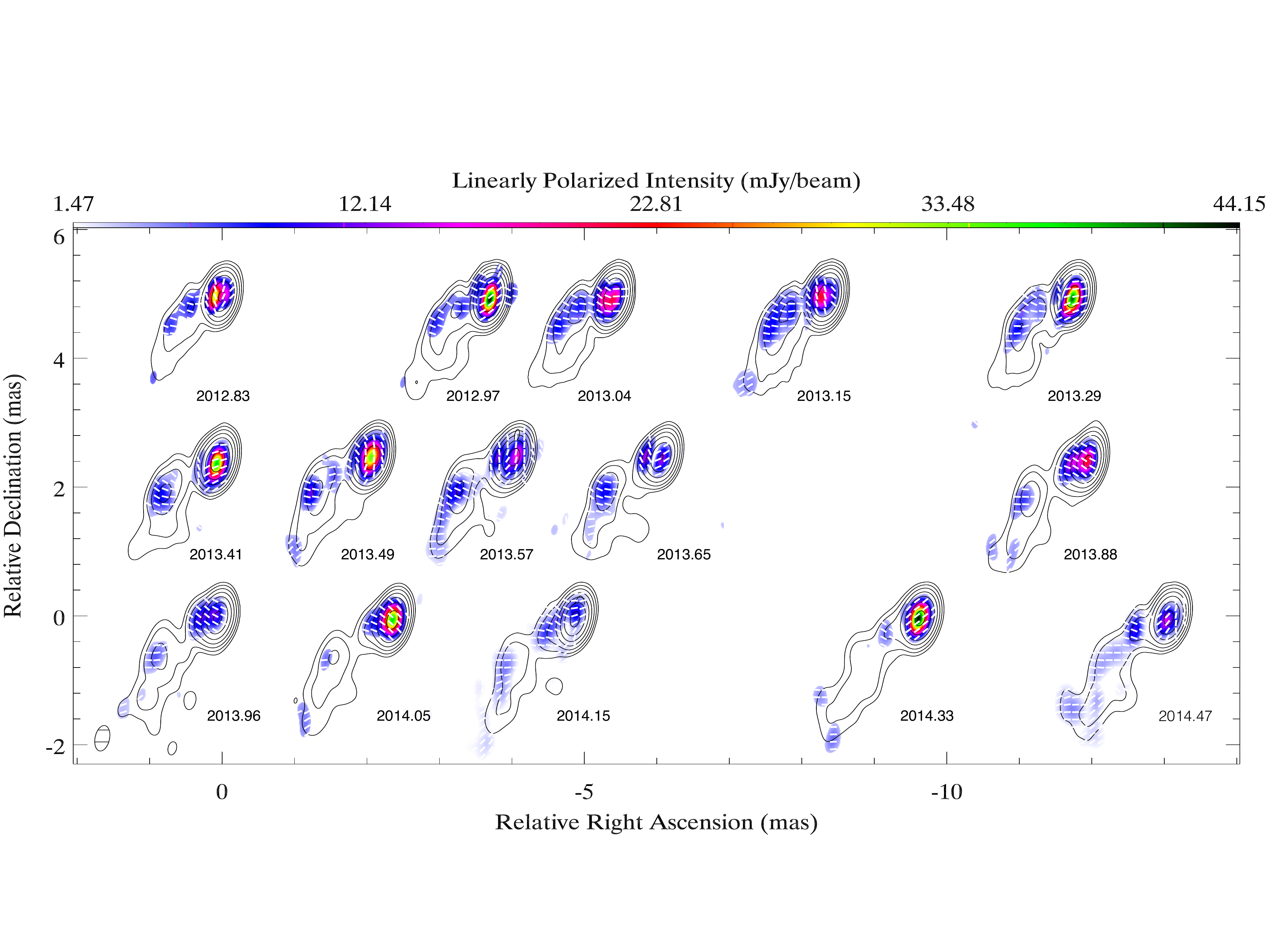}
\caption{Same as Figure~\ref{fig:CTA102_bw} with linearly polarized intensity in colors and white sticks symbolizing linear polarization angle.}
\label{fig:CTA102_pol}
\end{figure*}

\section{The parsec-scale jet}
\label{Sec:4}
\subsection{Physical Parameters of Components}

  VLBA images of CTA~102 at some selected epochs are displayed in Figures~\ref{fig:CTA102_28}-\ref{fig:CTA102_pol}. To carry out an analysis of the jet kinematics and flux density variability, we have fit with {\tt Difmap} the complex visibilities with a model source consisting of components described by circular Gaussian brightness distributions. For each epoch, we obtained a model-fit that provides information about the flux density (S), distance (r) and position angle ($\Theta$) relative to the core, and FWHM size (a) of each component. The core (labeled C0), considered stationary over the entire period, is identified with the unresolved component in the north-western (upstream) end of the jet. It is the brightest feature in the jet at most of the epochs. Polarization information has been obtained with an IDL program that calculates the mean values of the degree of polarization ({\it m}) and electric vector position angle (EVPA, $\chi$) over the image area defined by the FWHM size of each component. The uncertainties of both {\it m} and $\chi$ correspond to the standard deviations of their respective distributions. Model-fit parameters for all components and epochs are reported in Table~\ref{tabVLBA}.

  The accuracy of the model-fit parameters for each component depends on its brightness temperature, so that smaller uncertainties are expected for more compact components and higher flux densities. We have therefore established a criterion for quantifying the errors in the model fit parameters that is directly related to the observed brightness temperature, $T_b =7.5\times10^8 S/a^2 $ \citep[e.g.,][]{Jorstad:2005fk}, where $T_b$ is measured in Kelvins (K), {\it S} in Janskys (Jy), and {\it a} in milliarcseconds (mas).
  
  First, we select a representative sample of epochs and components with a wide range of {\it S}, {\it a} and {\it r} values. For each one of these components, we compute the error in the fitted parameters by analyzing how the reduced $\chi^2$ of the fit and resulting residual map change when varying the fitted parameters one at a time. We set a limit on the maximum allowed variation of the reduced $\chi^2$ of 20\%, corresponding to an increase by a factor of $\sim$1.5 in the peak levels of the residual map. According to this criterion, we assign a series of uncertainties in position and flux density to the components in the sample. We then relate the derived uncertainties with the measured brightness temperature, obtaining the following relations:
\begin{equation}
  {\sigma}_{xy} \approx 1.3\times10^4 \, T_b^{-0.6},
\end{equation}
\begin{equation}
  {\sigma}_S \approx 0.09 \, T_b^{-0.1},
\end{equation}
where $\sigma_{xy}$ and $\sigma_{S}$ are the uncertainties in the position (right ascension or declination) and flux density, respectively. These relations have been used to compute the errors in the position and flux density for all of the fitted components. To account for the errors in the flux calibration, we have added in quadrature a 5\% error to the uncertainty in flux density. The uncertainties in the sizes of components are also expected to depend on their brightness temperatures. Following \cite{Jorstad:2005fk}, we have assigned a 5$\%$ error to the sizes of the majority of components (those with flux densities $\ge$ 50 mJy and sizes of 0.1-0.3 mas) and a 10$\%$ error for more diffuse components.

\begin{table*}[htbp]
  \begin{center}
  \caption{VLBA 43 GHz Model-fit Components' Parameters}
  \begin{tabular}{c c c c c c c c} 
\hline
Epoch & Epoch & Flux  & Distance from & Pos. Angle & Major & Degree of & EVPAs \\ [0.07cm]
(year) & (MJD) & (mJy) & C0 (mas) & ($^{\circ}$) & axis (mas) & polarization ($\%$) & ($^{\circ}$)\\[0.07cm]
\hline
&&&& Component C0 &&&\\
\hline
2007.45 & 54264.5 & 3086 $\pm$ 313 & -- & -- & 0.017 $\pm$ 0.001 & -- & -- \\ [0.07cm] 
2007.53 & 54294.5 & 3423 $\pm$ 347 & -- & -- & 0.034 $\pm$ 0.002 & 1.5 $\pm$ 0.2 & 71.2 $\pm$ 5.7 \\ [0.07cm] 
2007.59 & 54318.5 & 2340 $\pm$ 239 & -- & -- & 0.038 $\pm$ 0.002 & 2.4 $\pm$ 0.1 & 76.4 $\pm$ 5.4 \\ [0.07cm] 
2007.66 & 54342.5 & 3163 $\pm$ 321 & -- & -- & 0.045 $\pm$ 0.002 & 1.6 $\pm$ 0.1 & 71.9 $\pm$ 6.6 \\ [0.07cm] 
2007.74 & 54372.5 & 2743 $\pm$ 279 & -- & -- & 0.045 $\pm$ 0.002 & 1.4 $\pm$ 0.1 & 82.9 $\pm$ 8.4 \\ [0.07cm] 
.\\
.\\
\hline
&&&& Component C1 &&&\\
\hline
2007.66 & 54342.5 & 296 $\pm$ 37 & 0.07 $\pm$ 0.01 & 119.2 $\pm$ 3.2 & 0.062 $\pm$ 0.003 & 1.6 $\pm$ 0.1 & 65.5 $\pm$ 7.1 \\ [0.07cm] 
2007.74 & 54372.5 & 307 $\pm$ 37 & 0.09 $\pm$ 0.01 & 104.4 $\pm$ 4.8 & 0.031 $\pm$ 0.002 & 2.2 $\pm$ 0.1 & -80.8 $\pm$ 6.3 \\ [0.07cm] 
2007.83 & 54405.5 & 168 $\pm$ 25 & 0.13 $\pm$ 0.02 & 119.5 $\pm$ 3.2 & 0.116 $\pm$ 0.006 & 2.6 $\pm$ 0.2 & 79.3 $\pm$ 5.9 \\ [0.07cm] 
2008.04 & 54482.5 & 575 $\pm$ 64 & 0.08 $\pm$ 0.01 & 146.2 $\pm$ 1.9 & 0.070 $\pm$ 0.003 & 0.7 $\pm$ 0.1 & 37.8 $\pm$ 5.5 \\ [0.07cm] 
2008.16 & 54524.5 & 894 $\pm$ 96 & 0.09 $\pm$ 0.01 & -166.3 $\pm$ 7.4 & 0.062 $\pm$ 0.003 & 0.5 $\pm$ 0.1 & 38.2 $\pm$ 5.6 \\ [0.07cm] 
.\\
.\\
\hline
 \multicolumn{8}{c}{\footnotesize (This table is available in its entirety in machine-readable form.)}
   \end{tabular}
  \label{tabVLBA}
  \end{center}
\end{table*}

\begin{table}[htbp]
  \begin{center}  
  \caption{Kinematics of Moving Jet Features}  
   \begin{tabular}{l c c c c} 
     \hline
     \hline
Name & N.Epoch &$\mu$ & $\beta_{app}$ & $T_{ej}$ \\ 
  && (mas/yr) & (c) & (year)\\ 
\hline
 N1 & 26 & 0.27$\pm$0.01 & 14.9$\pm$0.2 & 2009.12$\pm$0.02\\
 N2 & 18 & 0.35$\pm$0.01 & 19.4$\pm$0.8 & 2010.65$\pm$0.07 \\ 
 N3 & 10 & 0.49$\pm$0.03 & 26.9$\pm$1.8 & 2011.96$\pm$0.07 \\
 N4 & 6 & 0.21$\pm$0.02 & 11.3$\pm$1.2 & 2012.49$\pm$0.11 \\
\hline
   \end{tabular} 
   \label{table_1}
  \end{center}
\end{table}

\begin{figure}[htpb]
\centering
\includegraphics[width=0.5\textwidth]{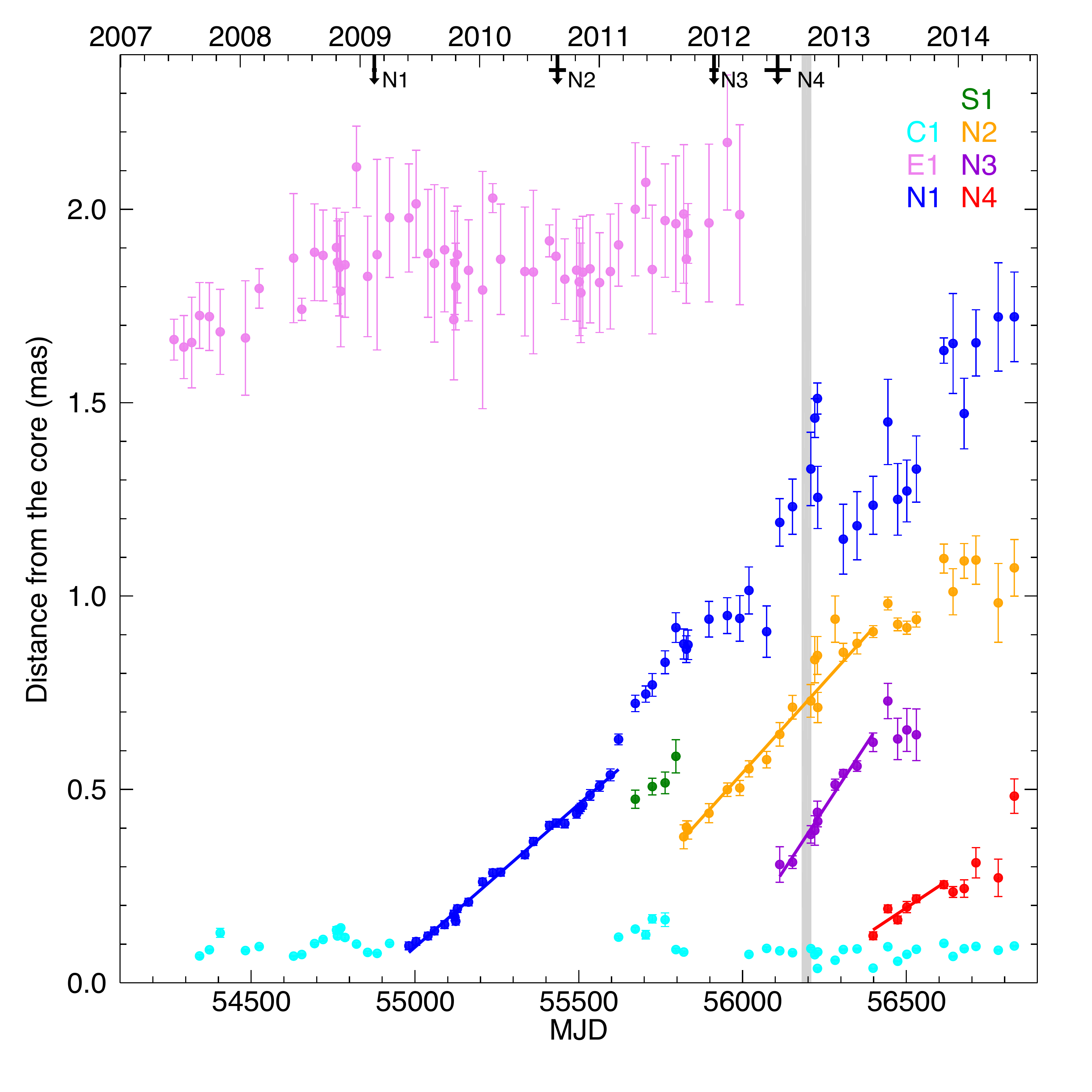}
\caption{Distance from the core vs. time for the 43 GHz model-fit components, with linear fits overlaid. Downward black arrows mark the time of ejection of each component with the respective error bar. The gray vertical stripe indicates the epoch of the $\gamma$-ray flare.}
\label{fig:Fit_Comp}
\end{figure}

\begin{figure}[htpb]
\centering
\includegraphics[width=0.5\textwidth]{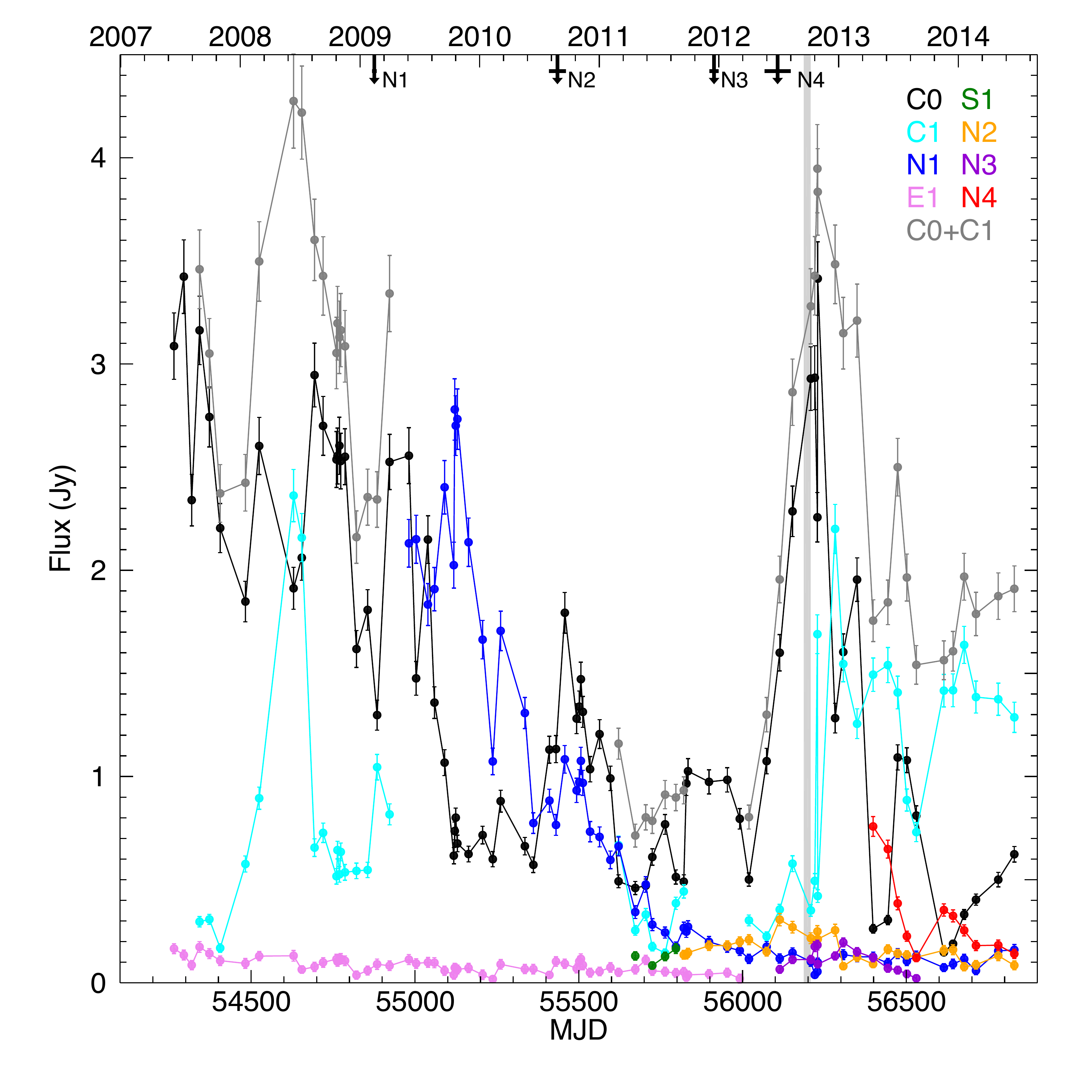} 
\caption{Light curves of 43 GHz model-fit components. Downward arrows and the gray vertical stripe indicate the same as in Figure~\ref{fig:Fit_Comp}.}
\label{fig:Flux_Comp}
\end{figure}

  Plots of separation and flux density versus time for the model-fit jet components, including the core, are presented in Figures~\ref{fig:Fit_Comp} and \ref{fig:Flux_Comp}. Besides the core, we have identified seven main components that could be traced reliably across multiple epochs. Component E1, located at $\sim$2 mas from the core, is a weak and extended feature that appears to be quasi-stationary across some epochs, or to move with a significantly slower velocity than other moving components (see Figure~\ref{fig:Fit_Comp}). A stronger and more compact component, C1, can be distinguished from the core at most of the observed epochs, located at a mean distance $r\sim$0.1 mas. Both quasi-stationary features have been observed previously by \cite{Jorstad:2005fk} and \cite{Fromm:2013uq, Fromm:2013fk}, and interpreted as recollimation shocks in the jet. We identify five other moving components, N1, N2, N3, N4 and S1. 
  Component S1 seems, however, to have a different nature: it appears to form in the wake of component N1 at a distance of $\sim$0.5 mas, and it is observed over only four epochs afterwards. Its properties are similar to those expected and observed previously for trailing components (Agudo et al. 2001; G\'omez et al. 2001; Jorstad et al. 2005). Linear fits of separation versus time have been obtained for the other moving components, N1, N2, N3, and N4, based on only those initial epochs at which an accurate position is obtained (see also Figure~\ref{fig:Fit_Comp}). 
  This yields the estimates for the apparent velocities and times of ejection (epoch at which the component coincides with the core) listed in Table~\ref{table_1}.

  Since we cannot directly measure the radial velocities of the jet features, a common approach to disentangle the contributions of the component's Lorentz factor and viewing angle in the observed proper motion is the use of the flux variability (e.g., Jorstad et al. 2005; Hovatta et al. 2009). Following Jorstad et al. (2005), we use causality arguments to infer the variability Doppler factor 
\begin{equation}
  \delta_{\mathrm{var}}={sD_{L}\over c\triangle t_{\mathrm{var}}(1+z)},
\end{equation}
where {\it s} is the disk-equivalent angular diameter (where s=1.6{\it a} for a Gaussian component fit with FWHM={\it a} measured at the epoch of maximum flux), and $D_{L}$ is the luminosity distance. The variability timescale is defined as $\triangle t_{\mathrm{var}}$=d{\it t}/ln($S_{\mathrm{max}}/S_{\mathrm{min}}$), where $S_{\mathrm{max}}$ and $S_{\mathrm{min}}$ are the measured maximum and minimum flux densities, respectively, and d{\it t} is the time in years between $S_{\mathrm{max}}$ and $S_{\mathrm{min}}$ \citep{Burbidge:1974ly}. This definition of $\delta_{\mathrm{var}}$ is valid under the assumption that the flux density variability timescale corresponds to the light-travel time across the component, which is valid as long as the radiative cooling time is shorter than the light crossing time and expansion time. Combining the estimated value of $\delta_{\mathrm{var}}$ with the measured apparent velocity, $\beta_{\mathrm{app}}=\beta \sin\theta/(1-\beta \cos\theta)$, where $\theta$ and $\beta$ are the viewing angle and velocity (in units of the speed of light) of the component, we can calculate the variability Lorentz factor, $\Gamma_{\mathrm{var}}$, and viewing angle, $\theta_{\mathrm{var}}$, using (Hovatta et al. 2009)
\begin{equation}
  \Gamma_{\mathrm{var}}={\beta_{\mathrm{app}}^{2}+\delta_{\mathrm{var}}^{2}+1\over 2\delta_{\mathrm{var}}}
\label{gamma}
\end{equation}
and
\begin{equation}
  \theta_{\mathrm{var}}=\arctan{2\beta_{\mathrm{app}}\over \beta_{\mathrm{app}}^{2} + \delta_{\mathrm{var}}^{2}-1}.
\label{theta}
\end{equation}
 
  Physical parameters of the moving components obtained from this method are reported in Table \ref{table_2}.
  
\begin{table}[htbp]
\begin{center}
\caption{Physical Parameters of Moving Jet Features}
\begin{tabular}{l c c c c c} 
 \hline
Name & $\triangle t_{var}$ & $a_{max}$\footnote{FWHM of the model-fit component calculated at the epoch of maximum flux} & $\delta_{var}$ & $\theta_{var}$ & $\Gamma_{var}$\\
 & (yr)  & (mas) & & ($^{\circ} $) & \\ 
 \hline
 N1 & 0.70 & 0.14 & 14.6 & 3.9 & 14.9 \\
 N2 & 1.12 & 0.33 & 22.4 & 2.5 & 19.6 \\
 N3 & 0.28 & 0.09 & 26.1 & 2.2 & 26.2 \\
 N4 & 0.20 & 0.08 & 30.3 & 1.2 & 17.3 \\
 \hline
\end{tabular}
\label{table_2}
\end{center}
\end{table}


\subsection{Kinematics and Flux Density Variability}

  By inspecting the light curves in Figure~\ref{fig:Flux_Comp} we can identify two flaring periods in the core: a prolonged first event that extends from mid-2007 to the beginning of 2009, and a second one between mid-2012 and the beginning of 2013, in coincidence with the main $\gamma$-ray flare.

  The peak flux of the first flare occurs between 2008 June and July, when both components C0 and C1 increase their flux densities, reaching a combined value of $\sim$ 4.2 Jy. Due to the proximity of C1 to the core, it is not always possible for the model fitting routine to clearly distinguish the two components, leading to high uncertainties in the flux ratio of the two features, as well as uncertainties in the position of C1. Because of this, Figure~\ref{fig:Flux_Comp} also shows the combined flux density of the core and component C1, providing the data needed to follow the total flux density within the core region of CTA~102.

 \begin{figure*}
\centering
\includegraphics[width=0.38\textwidth]{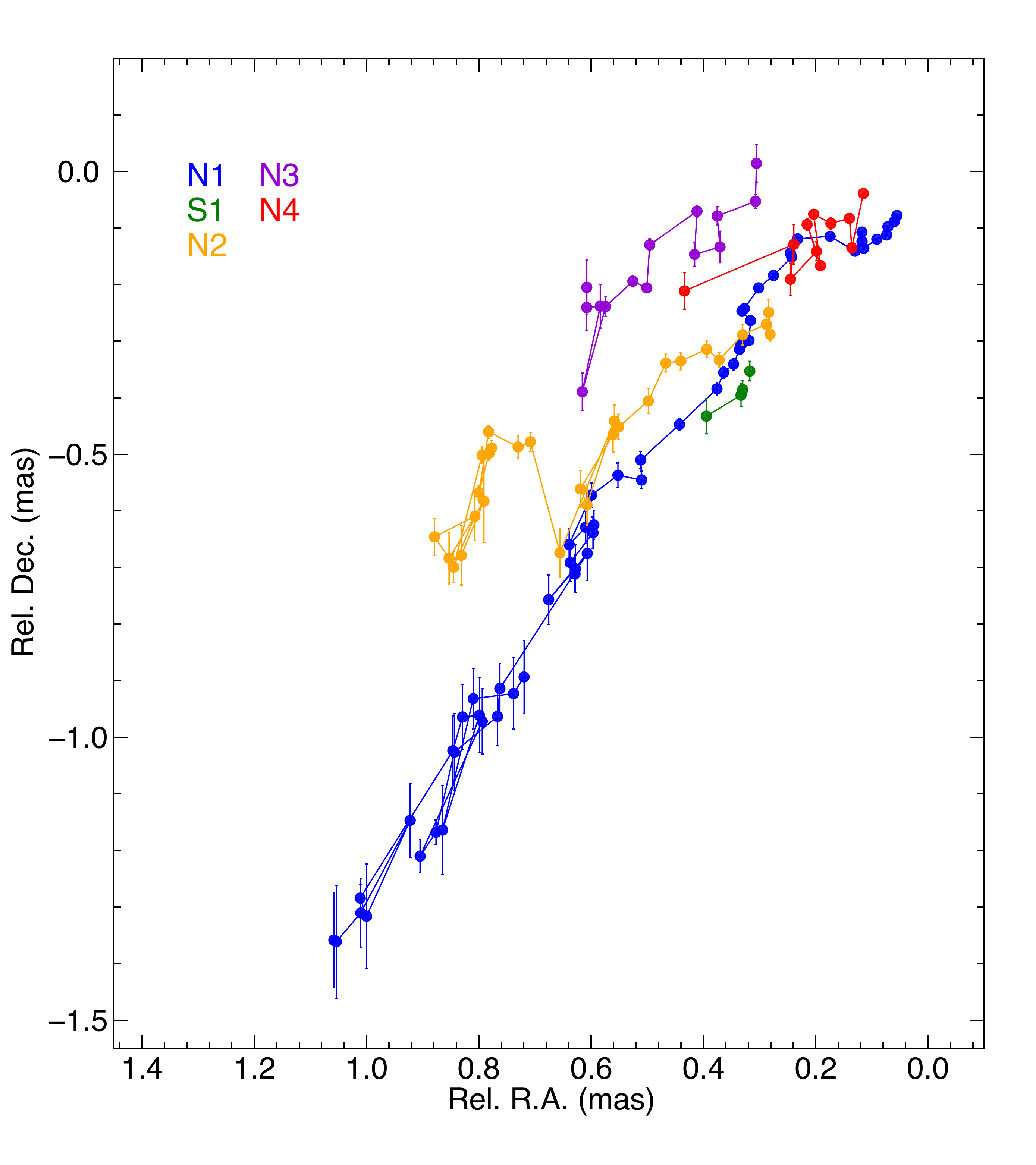}
\includegraphics[width=0.56\textwidth]{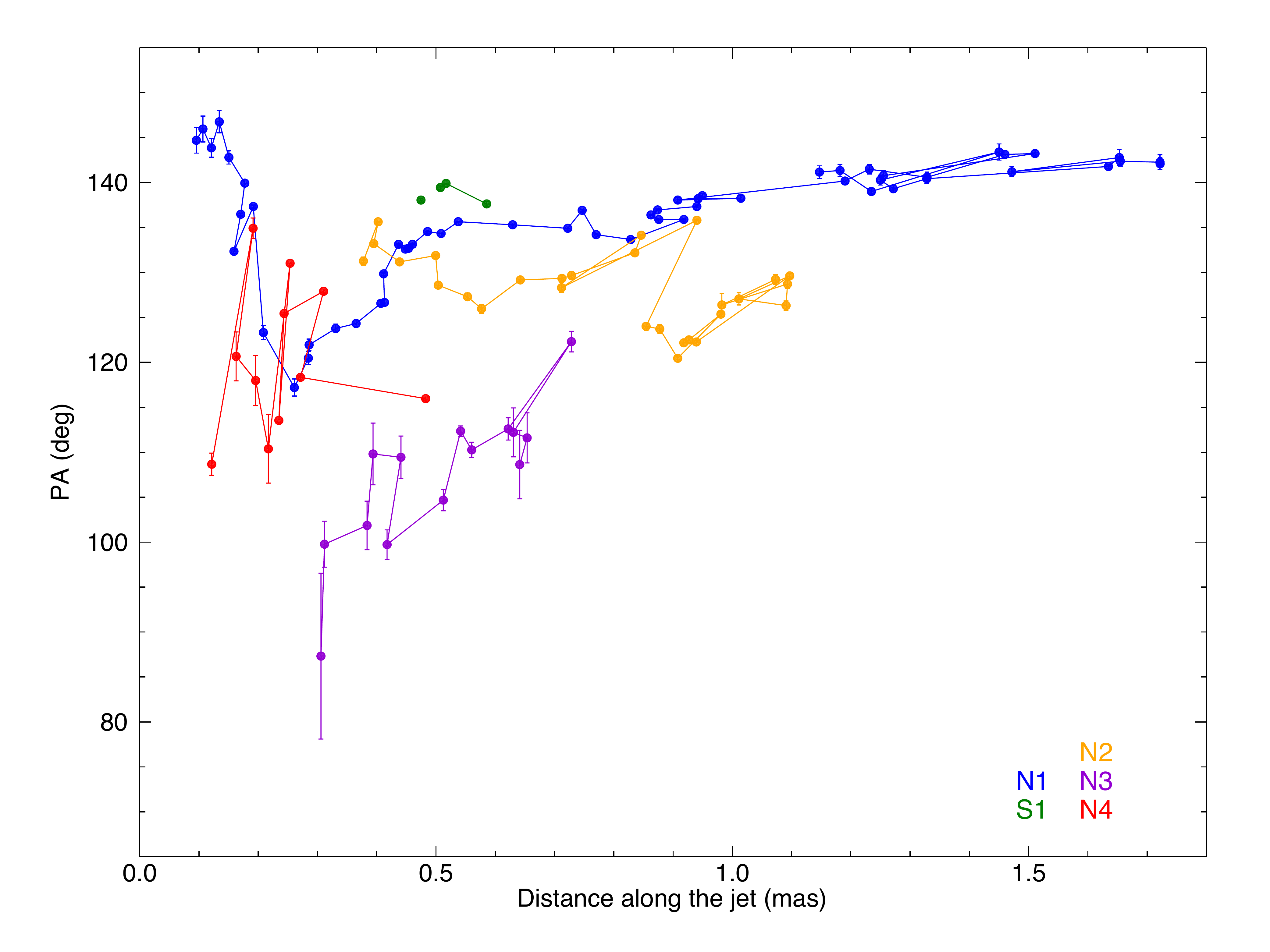} 
\caption{Trajectories ({\it left panel}) and position angles as a function of distance from the core ({\it right panel}) of the moving components in the jet.}
\label{fig:alfa_delta}
\end{figure*}
 
 \begin{figure}[htpb]
\centering
\includegraphics[width=0.9\textwidth]{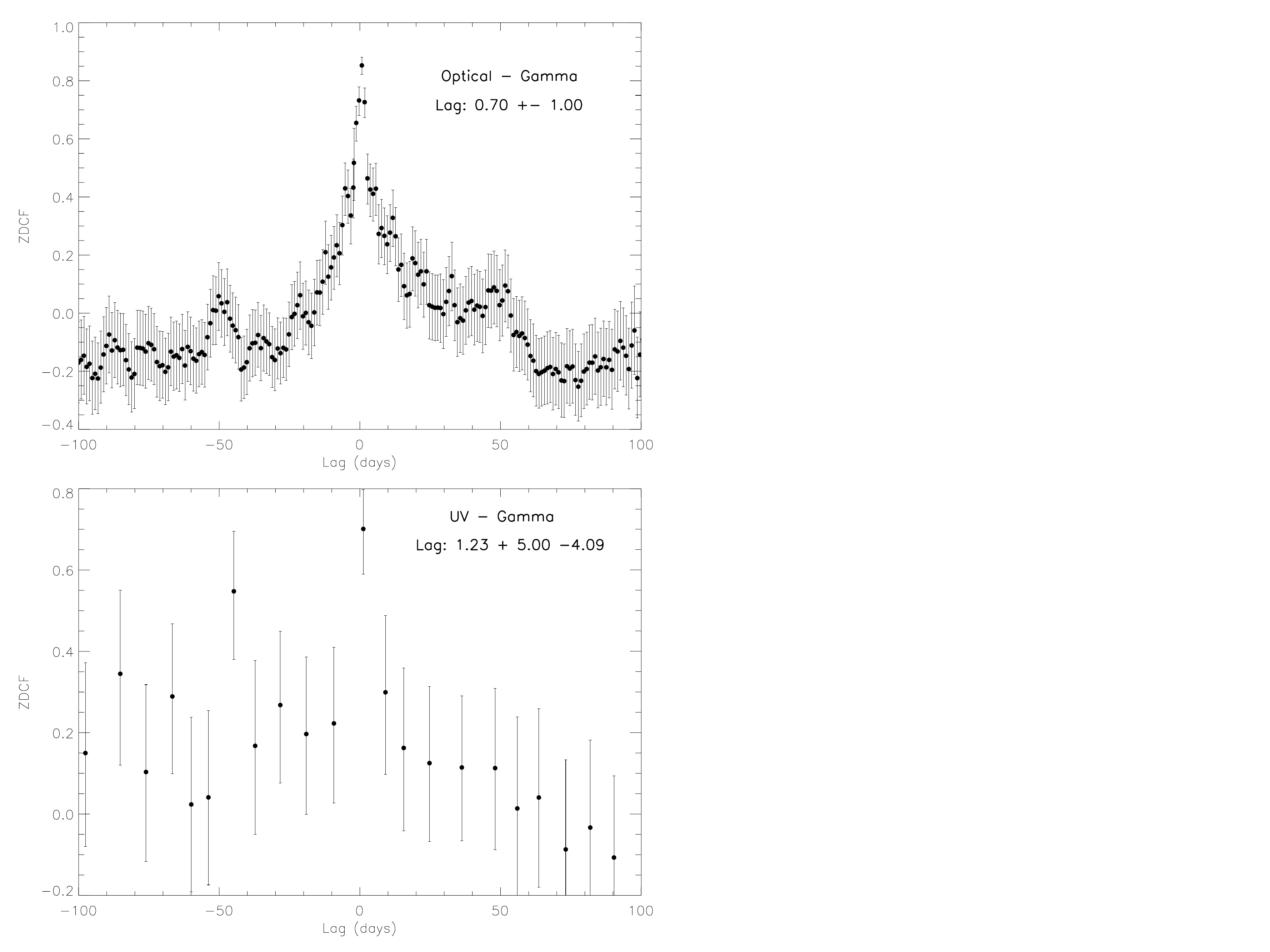}
\caption{Z-transformed discrete correlation function between optical and $\gamma$-ray data (upper panel) and UV and $\gamma$-ray data (lower panel). In each panel we report the time lag corresponding to the correlation peak with its respective 1-$\sigma$ error \cite[see][for more details]{Alexander:2013fk}.}
\label{fig:DCF}
\end{figure}

  The second flare in the mm-wave core begins in mid-2012, reaching its peak flux density at the end of 2012 October, close to the $\gamma$-ray flare (see Figure~\ref{fig:Flux_Comp}; section \ref{Sec:CC}). After the peak, the core region (C0 plus C1) remains in a high flux state until the last observing epoch, with a combined flux density oscillating around $\sim$ 2 Jy.
   
 Both flares in the core region are associated with the appearance of subsequent superluminal components. In the case of the first such flare, component N1 appears as a bright and well-defined feature that moves along the jet at $14.9\pm0.2\,c$ (see Figs.~\ref{fig:Fit_Comp}-\ref{fig:Flux_Comp} and Table \ref{table_1}). We also note that the ejection of component N1 corresponds to a change in the innermost structure of the compact jet, after which component C1 is no longer detected for almost two years. The second core flare leads to the ejection of component N4, which is significantly weaker than component N1 and has the slowest proper motion of the analyzed components (see Figs.\ref{fig:Fit_Comp}-\ref{fig:Flux_Comp} and Table \ref{table_1}).

  The values of the variability Doppler factors listed in Table \ref{table_2} correspond to a progressive increase with time, from 14.6 for component N1 to 30.3 for component N4. Previous estimations of the variability Doppler factor in CTA~102 range between 15.6 \citep{Hovatta:2009fk} and $22.3\pm4.5$ \citep{Jorstad:2005fk}, making N4 the superluminal knot with the highest Doppler factor to date. According to our analysis this unusually large value is due to a progressive re-orientation in the direction of ejection of knots, from $\theta_{\mathrm{var}}=3.9^{\circ}$ for component N1 to $\theta_{\mathrm{var}}=1.2^{\circ}$ for N4, which travels almost directly along the line of sight. This change in the jet orientation is readily apparent when analyzing each component's position angle shortly after the time of ejection, as well as their subsequent trajectories, as shown in Figure~\ref{fig:alfa_delta}.
  
  This smaller viewing angle of the jet with respect to the observer during the second radio flare, which appears to last until the end of our VLBA dataset (June 2014), is also in agreement with the significant differences observed between the ejections of components N1 and N4. While component N1 is clearly identified in the jet as a bright ($\gtrsim 2$ Jy) component soon after its ejection, most of the increase in the total flux density during the second radio flare appears to be associated with the core region (C0+C1), with component N4 representing only a small fraction of the flare. The smaller viewing angle of the jet also leads to a more difficult identification of component N4, which is not clearly discerned from the core until May 2014 (2014.33). Further support for the re-orientation of the jet toward the observed is also obtained from the analysis of the polarization, discussed in section \ref{Sec:Pol}.


\section{Cross-Correlation Analysis}
\label{Sec:CC}

To quantify the relationship among the light curves at the different wave bands, we perform a discrete cross-correlation analysis. The z-transformed discrete correlation function (ZDCF) described by \cite{Alexander:1997fk} has been designed for unevenly sampled light curves, as in our case. We use the publicly available {\it zdcf$\_$v1.2} and {\it plike$\_$v4.0.f90} programs\footnote{http://www.weizmann.ac.il/weizsites/tal/research/software/}, with a minimum number of 11 points inside each bin, as recommended for a meaningful statistical interpretation. We compute the DCF between each pair of light curves, including data from 100 days before to 100 days after the main $\gamma$-ray outburst. Time sampling of the light curves ranges from one day for the $\gamma$-ray data to tens of days in the case of some other wavebands (see Figure~\ref{fig:CTA102_mwl_lc}).

    Figure \ref{fig:DCF} displays the ZDCF analysis for the optical-$\gamma$ (upper panel) and UV-$\gamma$ (lower panel) data. We find that the correlation peaks between the $\gamma$-ray light curve and the optical and UV light curves give a time lag of $0.70\pm1$ and $1.23^{+5.00}_{-4.09}$ days, respectively, where a positive lag means that the $\gamma$-ray variations lead. We therefore conclude that the variations at the three wavelengths are essentially coincident within the uncertainties. 

  The sparser sampling of the X-ray data, as well as its double-peaked structure, precludes a reliable ZDCF analysis. However, we note that the first X-ray data peak is coincident with the $\gamma$-ray flare, and the second brighter X-ray flare occurs $\sim$50 days later (see Figure~\ref{fig:CTA102_mwl_lc}). 
  The triple-flare structure of the NIR light curve during the $\gamma$-ray flare (see Figure~\ref{fig:CTA102_NIR}) also prevents a unique interpretation of a cross-correlation analysis. Nevertheless, from inspection of the light curves, we see that the first, brightest peak in the NIR light curve (56193 MJD) is simultaneous with the $\gamma$-ray outburst within an uncertainty of 1 day, corresponding to the time sampling of both light curves.
  
  We obtain no significant correlation between the millimeter-wave and $\gamma$-ray light curves. This can be due to the different timescales associated with the emission at these wavebands, as suggested also for other blazars \citep[e.g., 1156+295,][]{Ramakrishnan:2014ys}. 
  The rise time for the millimeter wave band is of the order of months, while for the $\gamma$-rays it is of the order of few days. We note, however, that the 1, 3, and 7 mm light curves contain a significant flare coincident with the $\gamma$-ray 
  outburst.

\section{Polarized Emission}
\label{Sec:Pol}

\begin{figure}[htpb]
\centering
\includegraphics[width=0.48\textwidth]{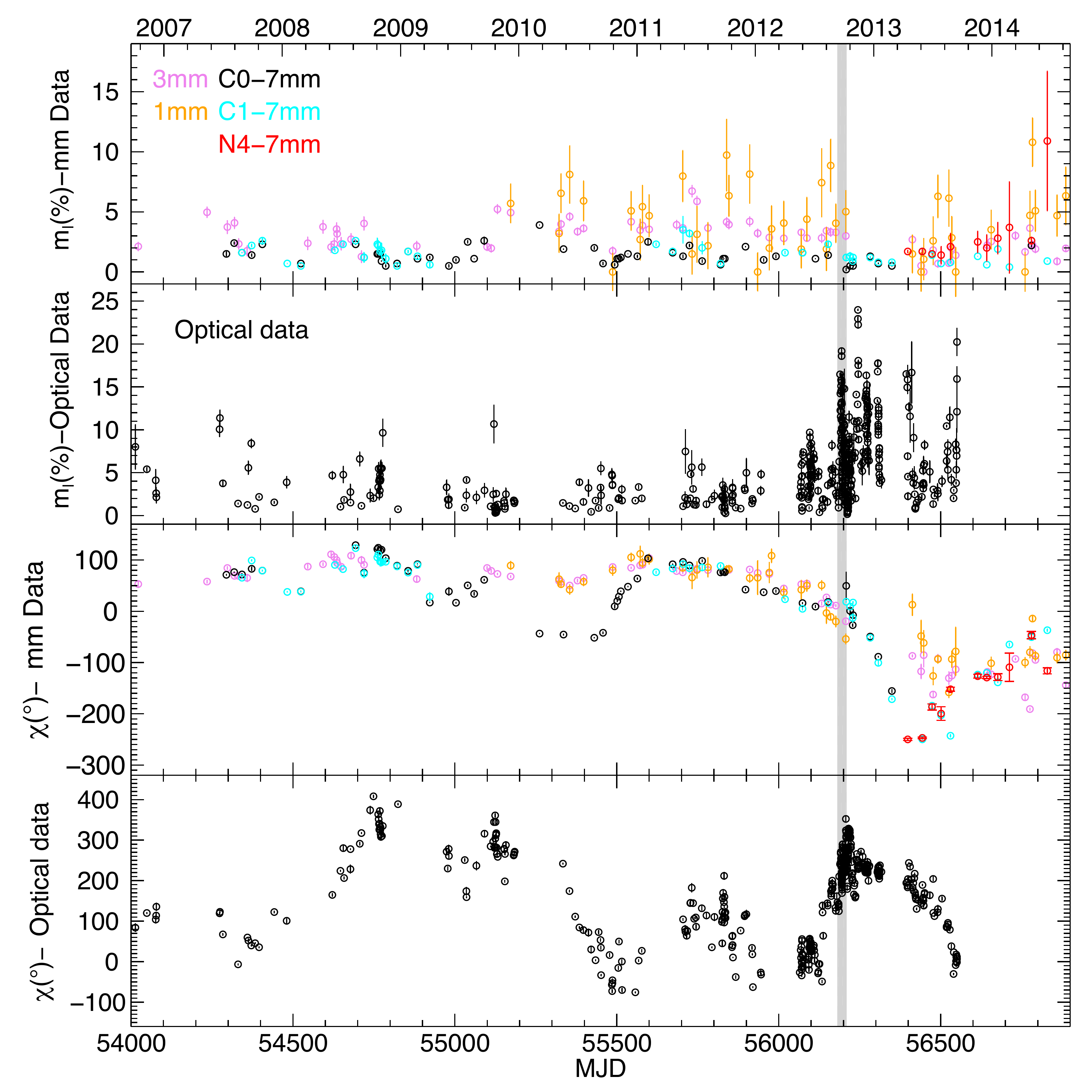}
\caption{Optical and millimeter-wave linear polarization over the period MJD 54000--56800. The first two panels display the degree of mm-wave and optical polarization, respectively. The third and fourth panels display the EVPAs at millimeter and optical wavelengths, respectively. The gray vertical stripe indicates the epoch of the main $\gamma$-ray flare.}
\label{fig:CTA102_mwl_pol}
\end{figure}

 Figure \ref{fig:CTA102_mwl_pol} shows the optical and millimeter-wave linear polarization between MJD 54000 and 56900, covering the period of the $\gamma$-ray flare. To solve for the $\pm$$n\pi$ ambiguity in the electric vector position angle (EVPA), we assume the slowest possible variation in time, applying a $\pm\pi$ rotation between two consecutive measurements when the magnitude of the EVPA change would otherwise exceed $\pi/2$.

  No significant increase in the degree of polarization at millimeter wavelengths is observed during the $\gamma$-ray flare, but the EVPAs display a progressive rotation starting about one year prior to the $\gamma$-ray flare. Figure~\ref{fig:CTA102_mwl_pol} shows that between 2007 and mid-2011 the EVPAs at 3 and 1~mm are distributed around a mean value of $\sim$100$^{\circ}$. After this, the polarization at millimeter wavelengths starts a slow rotation by almost 80$^{\circ}$ in one year (from 2011 July to 2012 August), until the flare epoch. In coincidence with the $\gamma$-ray flare, the rate of EVPA rotation in the VLBI core and stationary component C1 increases significantly, leading to a rotation of almost 200$^{\circ}$ in one year. Subsequently, component N4 appears and the EVPAs of both C1 and N4 rotate again toward values similar to those at 1 and 3 mm, reaching $\sim -100^{\circ}$. It is possible that during the flare, while the new superluminal component N4 is crossing the core zone, the EVPAs of the innermost region at 7 mm rotate due to the passage of the component. After this, when N4 can be distinguished from C1 and C0, the EVPAs at 7 mm again follow the general behavior of the EVPAs at shorter millimeter wavelengths. A similar discrepancy between the 1-3~mm and 7~mm EVPAs occurs between mid-2009 and mid-2010, when component N1 is ejected and becomes brighter than the core until mid-2010 (see Figure~\ref{fig:Flux_Comp}).

  \begin{figure}[htpb]
\centering
\includegraphics[width=0.5\textwidth]{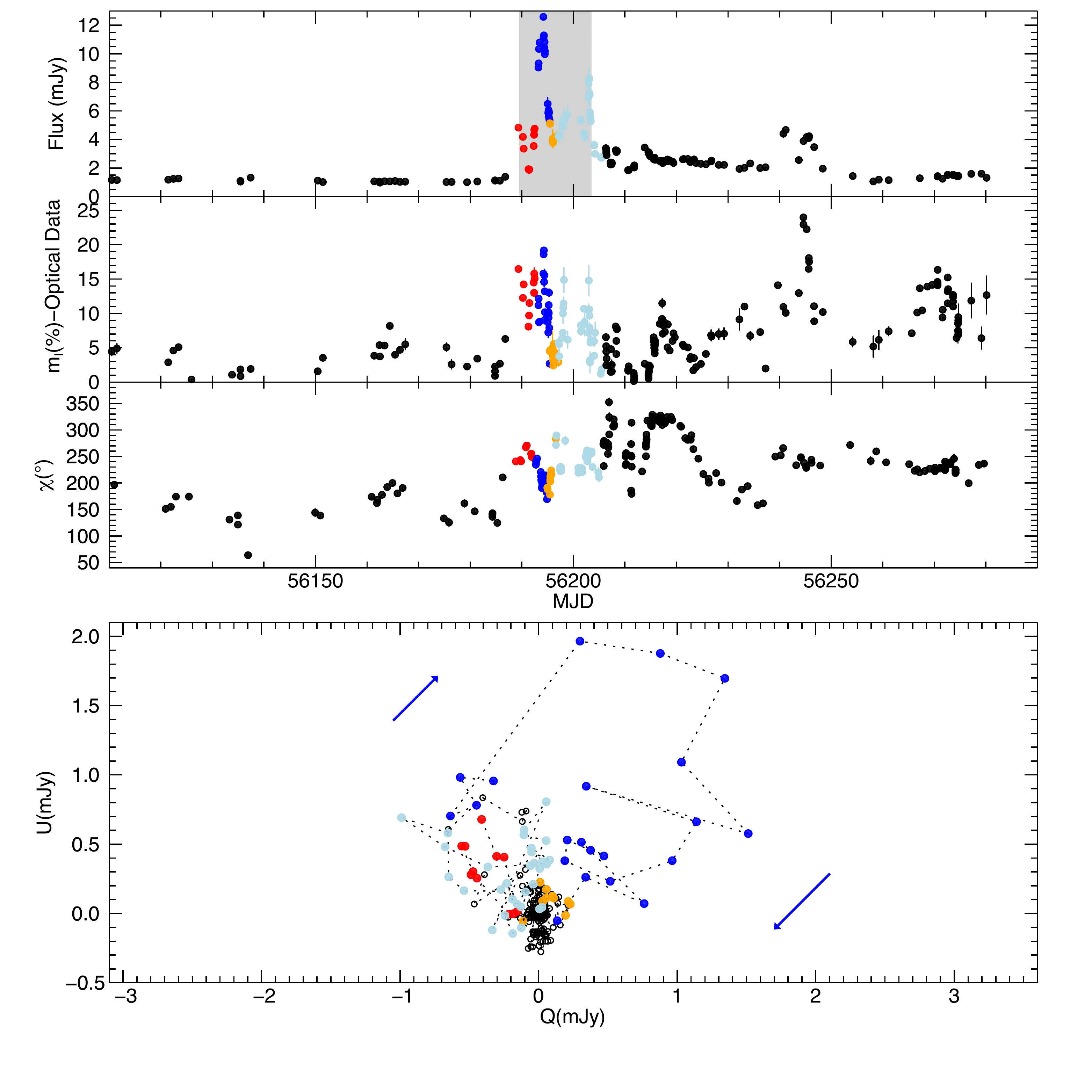}
\caption{The upper panel displays, from top to bottom, the light-curve, degree and time evolution of polarization, and EVPA at optical frequencies. Each colored mark corresponds to the period over which we plot the $U$ and $Q$ Stokes parameters in the lower panel. Blue points mark a clockwise rotation cycle that occurs in coincidence with the total intensity peak. The gray vertical stripe indicates the time range of the main gamma-ray flare.}
\label{fig:UvsQ}
\end{figure}

  We can then distinguish the rapid rotation of polarization vectors observed in the VLBI components at 7~mm after the flare from the slower rotation observed at 1 and 3~mm. The latter leads to a rotation of the mm-EVPAs of $\sim$200$^{\circ}$ over 3 years (from 2011 July to 2014 August).
  This progressive rotation in the EVPAs can be produced by a change in the orientation of the innermost jet, which would be in agreement with the larger Doppler factor and smaller viewing angle of component N4 associated with the $\gamma$-ray event, as discussed previously (see {\S}~\ref{Sec:4}).

  The optical polarization executes rapid and pronounced changes in both degree of polarization and EVPA associated with the $\gamma$-ray flare. Figure \ref{fig:UvsQ} displays an expanded view of the optical polarization data near the time of the $\gamma$-ray flare, with four different time ranges marked in different colors. Before the peak at optical frequencies on MJD 56194, the source undergoes a period of rapid changes in both total and polarized emission (marked in red), and the EVPAs rotate by almost $30^{\circ}$. The plot of Stokes parameters U vs.\ Q in Figure~\ref{fig:UvsQ} reveals a clear clockwise rotation of the EVPAs (marked in blue), in coincidence with the main flare in total flux and a rapid change in the degree of polarization. This clockwise rotation has been previously reported by \cite{Larionov:2013uq}.
  
  If we assume a model in which a relativistic shock does not cover the entire cross-section of the jet and is moving down the jet following helical magnetic field lines, which also propagate downstream, then we expect to observe a rotation in the EVPA. 
  This should be accompanied by a change in the degree of polarization, with a minimum in the middle of the rotation, where the flaring region contains magnetic field lines with opposite polarity \citep[e.g.,][]{Vlahakis2006,Marscher:2008vn, Larionov:2013kx}. Evidences of a helical magnetic field in CTA~102 jet can also be found in the detection of negative circular polarization \citep{Gabuzda:2008fk} and in a gradient in the rotation measure across the jet width at about 7 mas from the core \citep{Hovatta:2012fk}, both in MOJAVE observations.


\section{Spectral Energy Distribution}

\begin{figure}[htpbt]
\centering
\includegraphics[width=0.495\textwidth]{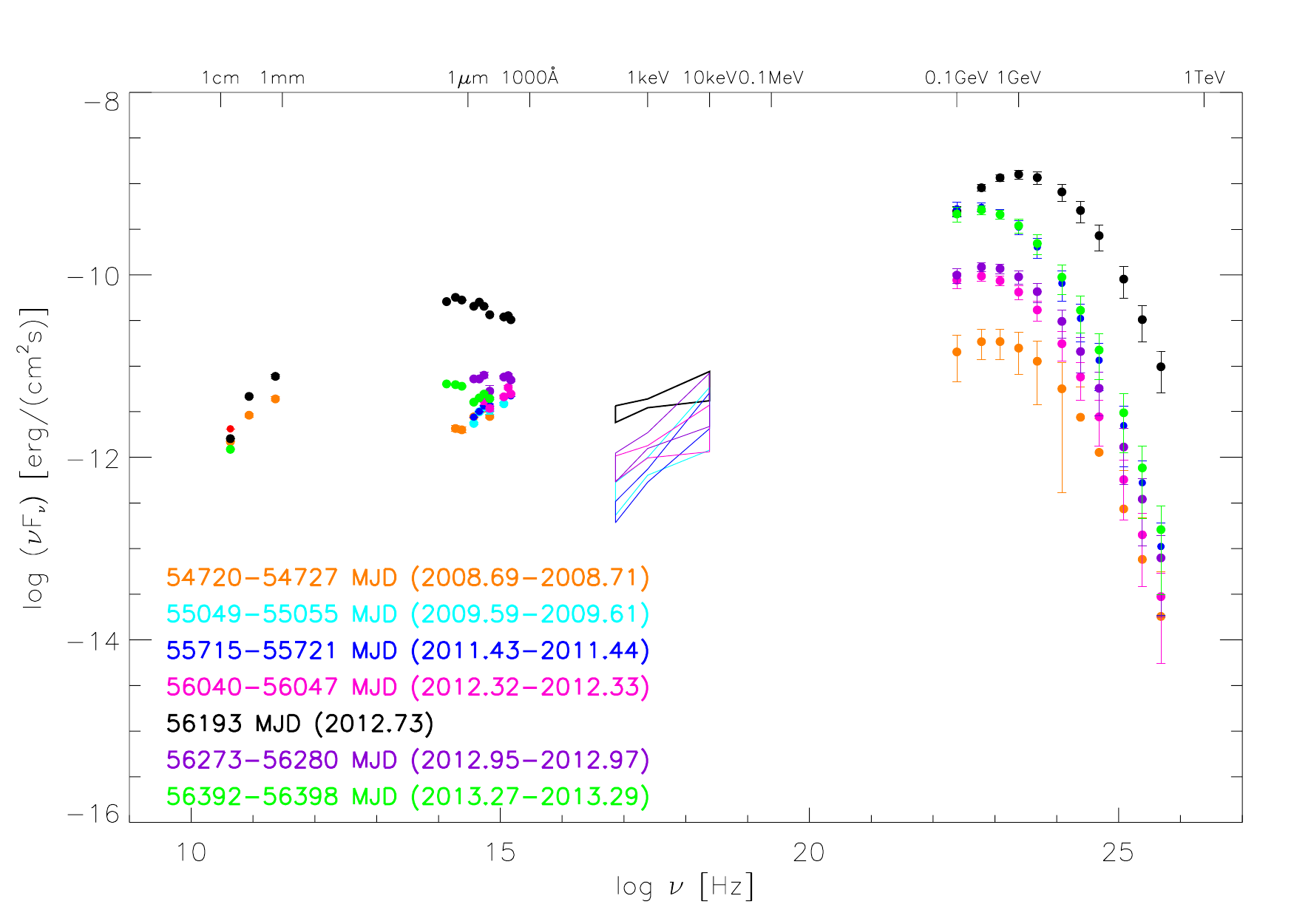}
\caption{Spectral energy distribution of CTA~102 from millimeter-wave to $\gamma$-ray frequencies during the brightest $\gamma$-ray flare (black) and at other observing epochs, as labeled in the figure and discussed in the text.}
\label{fig:SED}
\end{figure}

We have computed the spectral energy distribution (SED) of the source from millimeter to $\gamma$-ray wavelengths at several epochs (see Figure~\ref{fig:SED}): two epochs between MJD 54720 and MJD 55055, corresponding to $\gamma$-ray quiescent states; the epochs of the first $\gamma$-ray flare (MJD 55715--55721), the main flare (MJD 56193), and the third $\gamma$-ray flare (MJD 56392--56398); one epoch of a quiescent state between the first and second flares (MJD 56040--56047), and a second one between the second and third flares (MJD 56273--56280). For the main $\gamma$-ray flare, all data are simultaneous except for the mm-wave data, which corresponds to MJD 56208. For the other epochs, we have considered a range of time (as indicated in Figure~\ref{fig:SED}) in order to cover the entire energy range.

  By examining Figure \ref{fig:SED}, we observe that, during the multi-wavelength flare in 2012 (black points) both, the synchrotron and the inverse Compton peaks increased. The synchrotron peak frequency during the flare is close to 10$^{14}$ Hz, at the upper end of the frequency range 10$^{12}$-10$^{14}$ Hz of synchrotron peaks observed in luminous blazars \citep{Ghisellini:2008fk}. At the same time, we also observe a shift in the inverse Compton peak to higher frequencies, which leads to a hardening of the spectrum between 0.1 and 1 GeV. We note that neither of the other two, weaker $\gamma$-ray flares display a similar shift in the peak of the inverse Compton spectrum. In particular, the weaker first and third $\gamma$-ray flares display a very similar inverse Compton spectrum, peaking at nearly the same frequency, with only a minor hardening toward higher energies during the third flare. The shift of the $\gamma$-ray peak toward higher frequencies during the main $\gamma$-ray flare can be explained by a change in the viewing angle -- leading to an increase in Doppler factor -- of the emitting region, in support of our hypothesis of a reorientation of the jet toward the line of sight during the multi-wavelength flare (see {\S}{\S}~\ref{Sec:4} and \ref{Sec:Pol}).
    
  The ratio of the inverse Compton to synchrotron peaks is of the order of 10, which is not sufficiently large to rule out synchrotron self-Compton scattering as the main mechanism for the production of the $\gamma$-ray emission \citep[e.g.,][]{Sikora:2009uq}.
 
 
\section{Discussion and results}
  We have presented a multi-wavelength polarimetric study of the quasar CTA~102 during an unprecedented $\gamma$-ray outburst that was observed between 2012 September 23 and October 2. We find that the $\gamma$-ray outburst occurred simultaneously with flares from millimeter to X-ray wavelengths, with the exception that the 1 and 7 mm light curves peak almost one month after the shorter-wavelength flares. However, all of the millimeter-wave light curves begin to increase before the shorter-wavelength outburst, but took longer times to reach maximum flux and then to decay. 

  Our DCF analysis confirms the coincidence between the $\gamma$-ray flare and the optical-UV flare. The same analysis does not provide unambiguous correlation between the X-ray or NIR and the $\gamma$-ray light curves because of the multi-peak structure of the flare at these two frequencies and the relatively sparse sampling.

  The {\it Fermi} LAT daily light curve reveals two more $\gamma$-ray outbursts apart from that in 2012: one in 2011 June and the other in 2013 April. Both outbursts are weaker and ``orphan''. Only the bright outburst in 2012 September-October is coincident with flares at the other wavelengths and with the emergence of a new superluminal knot from the radio core.

  We have combined our multi-wavelength study of light curves with an analysis of multi-epoch VLBA observations at 43 GHz that provide the necessary angular resolution to follow the evolution of the jet during the outburst. In the 43 GHz VLBA images, we observe the ejection of multiple superluminal knots from the radio core during the analyzed period, but only one of these knots, N4, is associated with a $\gamma$-ray flare. Component N4 was ejected in $2012.49\pm0.11$, within a time range between 47 and 127 days before the main $\gamma$-ray flare in 2012 (2012.73), when the radio core started to increase in flux density. The interaction between a traveling feature and the stationary radio core appears to have triggered a number of $\gamma$-ray outbursts in blazars \citep[e.g.,][]{Ramakrishnan:2014ys, Morozova:2014uq} and radiogalaxies~\citep{Grandi:2012fk,Casadio:2015fk}. However, not every ejection of a new knot leads to a $\gamma$-ray flare. For instance, it is not clear why a $\gamma$-ray flare is associated with N4 in CTA~102 and not with the other moving radio components.
  
  From the analysis of model-fit components at 43 GHz, we deduce that the jet changed its orientation with respect to the observer when component N4 was ejected. This is derived from an analysis of the variability Doppler factor and viewing angle, which indicates that a progressive increase in the Doppler factor occurred, caused by a re-orientation of the jet toward the line of sight. This led to a minimum viewing angle of $\theta\sim$1.2$^{\circ}$ when component N4 was ejected during the $\gamma$-ray outburst. This change in the orientation of the jet is  supported by the observed progressive, slow rotation of the millimeter-wave EVPAs starting almost one year before the ejection of N4 and the $\gamma$-ray flare. We therefore conclude that the $\gamma$-ray emission in CTA~102 is related to a decrease in the viewing angle of the jet.
  
  This correlation between $\gamma$-ray activity and orientation of the jet has been already observed in other BL Lac objects \citep{Marscher:2008vn, Larionov:2010fk, Rani:2014vn}, quasars \citep{Abdo:2010uq,Raiteri:2011kx,Jorstad:2013uq}, and radio galaxies \citep{Casadio:2015fk}, although there are different interpretations regarding the cause of the change in orientation. Some authors consider a bent or precessing jet, while others suggest a helical jet with the radiating component following this helical path. A helical trajectory could also be the consequence of magnetic field lines twisting around a conical or parabolic jet \citep{Vlahakis2006}.

  In the case of CTA~102, there are indications of a helical magnetic field structure \citep{Gabuzda:2008fk, Hovatta:2012fk}. We associate the fast variability in the polarized optical emission, as well as the clockwise rotation displayed in the EVPAs during the outburst, with the helical path followed by the superluminal component in its motion along the outwardly propagating magnetic field lines. On the other hand, a number of similar rotations of the mm-wave and optical polarization vectors occurred in both the clockwise and counterclockwise directions over the entire 2004--2014 monitoring period. This can be interpreted in terms of random walks of a turbulent magnetic field \citep{Jones1988,DArcangelo2007,Marscher2014}. Early results from the RoboPol program show that, while many EVPA rotations related to $\gamma$-ray flares can be produced by a random walk process, some are not \citep{Blinov:2015kx}. If the rotation associated with a $\gamma$-ray flare is caused by a helical geometry of the magnetic field, then future such outbursts should be accompanied by similar clockwise rotations.
  
  The observed long-term rotation in millimeter-wave polarization vectors, together with the slower proper motion associated with component N4, suggest a change in the jet orientation, so that it becomes more closely aligned with the line of sight during the ejection of component N4 and the multi-wavelength flare. 

  The close timing of the $\gamma$-ray, X-ray, UV, and optical flares suggests co-spatiality of the emission at all these frequencies. Knot N4 was 0.025 to 0.07 mas downstream of the core when the $\gamma$-ray flare occurred, i.e., it had not yet reached feature C1 at $\sim$0.1 mas. This is confirmed by the increase in flux density in the 7 mm core during the $\gamma$-ray outburst. Hence we conclude that the bright $\gamma$-ray outburst occurred inside the mm-wave core region.

  We observe component N4 for the first time in the VLBA images on 2013 April (MJD 56398), when it was located at $r\sim$0.12 mas. The $\gamma$-ray flare in 2013 April occurred between MJD 56387 and 56394. Therefore, a possible interpretation of this flare is the passage of component N4 through C1, interpreted by \cite{Fromm:2013fk} as a possible recollimation shock.

   If the radio core were located within $\sim1$ pc of the black hole (BH), then the accretion disk or the broad line region could provide the necessary photon field to explain the high energy emission through external Compton scattering. The 43 GHz radio core in CTA~102 must be coincident with, or downstream of, the 86 GHz core that is located at a distance of $7.5\pm3.2$ pc \citep[$\sim8.5\times10^{4}$ gravitational radii for a BH mass of $\sim8.5\times10^{8} M_{\odot}$;][]{Zamaninasab:2014fk} from the BH \citep{Fromm:2015uq}. A similar scaled distance, $\sim10^4$--$10^5$ gravitational radii, has been determined also for two radio galaxies, 3C~111 and 3C~120~\citep{Marscher:2002kx, Chatterjee:2009ve, Chatterjee:2011qf} and two blazars, BL Lac and 3C~279~\citep{Marscher:2008vn, Abdo:2010uq}. For a mean viewing angle of the jet of CTA~102 of 2.6$^{\circ}$ \citep{Jorstad:2005fk,Fromm:2015uq}, the distance of N4 from C0 is 4.6--13 pc, hence the $\gamma$-ray outburst took place more than 12 pc from the BH. At this location, there should be a negligible contribution of photons from the disk or the broad line region, nor from the dusty torus \citep[located $\sim$1.6 pc from the BH;][]{Pacciani:2014fk}, for external Compton scattering to produce the high energy flare. The lack of a suitably strong external source of photons favors synchrotron self-Compton (SSC) scattering of NIR to UV photons by electrons in the jet with energies $\sim 10$ times the rest-mass energy, as the source of the $\gamma$-ray emission. The ratio of $\gamma$-ray to infrared (synchrotron) luminosity is $\lesssim10$, sufficiently low to be consistent with the SSC process.

\section{conclusion}
  Our study of the time variability of the multi-wavelength flux and linear polarization of the quasar CTA~102 confirms its erratic blazar nature, revealing both strong connections across wave bands in one outburst and no obvious connections for other events. The bright $\gamma$-ray outburst in late 2012 was accompanied by contemporaneous flares at longer wavelengths up to at least 8 mm, with the increase in mm-wave flux starting before the $\gamma$-ray activity. The polarization vector at both optical and millimeter wavelengths rotated from the time of the $\gamma$-ray peak until $\sim 150$ days later. A new superluminally moving knot, N4 --- the feature with the highest Doppler beaming factor during our monitoring, according to our analysis --- was coincident with the core in the 43 GHz VLBA images 47--127 days prior to the $\gamma$-ray peak. We conclude that the outburst was so luminous because the jet (or, at least, the portion of the jet where most of the emission occurs) had shifted to a direction closer to the line of sight than was previously the case. The time delay between the epoch when N4 crossed the centroid of the core (feature C0) and the epoch of peak $\gamma$-ray emission implies that the main flare took place $\gtrsim 12$ pc from the black hole. At this distance, the only plausible source of seed photons for inverse Compton scattering is NIR to UV emission from the jet itself. The ratio of $\gamma$-ray to infrared luminosity is only $\sim10$ at the peak of the outburst, low enough to be consistent with SSC high-energy emission.
  
  Multiple superluminal knots appeared in the jet during the 7 years covered by our VLBA observations. These include a very bright component (N1) ejected in $2009.12\pm0.02$ and associated with a significant mm-wave flare in the core region ($\sim$4.2 Jy). Yet only component N4 is related to a flare at $\gamma$-ray energies. Two strong ``orphan'' $\gamma$-ray flares have no apparent optical counterparts. A strong mm-wave event with neither a $\gamma$-ray nor optical counterpart can be explained by an inability of the event to accelerate electrons up to energies $\sim 10^4 mc^2$ needed to radiate at such frequencies, although the reason for this inability is unclear. Orphan $\gamma$-ray flares might be explained by a knot crossing a region where there is a higher local density of seed photons for inverse Compton scattering \citep{Marscher2010,MacDonald2015}. Indeed, the second orphan flare corresponds to the time of passage of knot N4 through stationary feature C1 (located ~$\sim 0.1$ mas from the core), which could be such a region. 
  
  During the multi-wavelength outburst we observe intra-day variability in the optical polarized emission, as well as a clockwise rotation in optical EVPAs. This rotation could be caused by a spiral path traced by the knot moving along helical magnetic field lines that propagate outwards relativistically. Alternatively, the various rotations of the polarization vector seen in our dataset, which are in the clockwise and counterclockwise direction over different time ranges, could be mainly random walks caused by a turbulent magnetic field.

  CTA~102 displays the complex behavior characteristic of the blazar class of active galactic nuclei. Nevertheless, we have found possible connections between variations in the multi-wavelength flux and polarization and in the structure of the jet in some events. Continued monitoring of CTA~102 and other bright blazars at multiple wave bands with as dense sampling as possible, combined with mm-wave VLBI imaging, can eventually determine which connections are robust and the extent to which stochastic processes dominate the behavior of blazars.

\acknowledgments
This research has been supported by the Spanish Ministry of Economy and Competitiveness (MINECO) grant AYA2013-40825-P. The research at Boston University (BU) was funded in part by NASA Fermi Guest Investigator grants NNX14AQ58G and NNX13AO99G, and Swift Guest Investigator grant NNX14AI96G. Iv\'an Agudo acknowledges support by a Ram\'on y Cajal grant of the 
MINECO. The VLBA is operated by the National Radio Astronomy Observatory. The National Radio Astronomy Observatory is a facility of the National Science Foundation operated under cooperative agreement by Associated Universities, Inc. The PRISM camera at Lowell Observatory was developed by K.\ Janes et al. at BU and Lowell Observatory, with funding from the NSF, BU, and Lowell Observatory.
St.Petersburg University team acknowledges support from Russian RFBR grant 15-02-00949 and St.Petersburg University research grant 
6.38.335.2015. This research was conducted in part using the Mimir instrument, jointly developed at Boston University and Lowell Observatory and supported by NASA, NSF, and the W.M. Keck Foundation. The Mimir observations were performed by Lauren Cashman, Jordan Montgomery, and Dan Clemens, all from Boston University. 
This research is partly based on data taken at the IRAM 30m Telescope. IRAM is supported by INSU/CNRS (France), MPG (Germany), and IGN (Spain).
The Submillimeter Array is a joint project between the Smithsonian Astrophysical Observatory and the Academia Sinica Institute of Astronomy and Astrophysics and is funded by the Smithsonian Institution and the Academia Sinica.
Data from the Steward Observatory spectropolarimetric monitoring
project were used. This program is supported by Fermi Guest Investigator
grants NNX08AW56G, NNX09AU10G, and NNX12AO93G. The Mets\"ahovi team acknowledges the support from the Academy of Finland
to our observing projects (numbers 212656, 210338, 121148, and others).


\end{document}